\documentclass[preprint,aps,onecolumn,floats,floatfix,amsmath,nofootinbib]{revtex4-1}
\usepackage{graphicx,array,dcolumn}
\usepackage{calc,tabularx, epsfig,mathrsfs}
\usepackage{hyperref}

\usepackage{tikz}
\usetikzlibrary{decorations.pathmorphing}
\usetikzlibrary{arrows}

\usepackage{amsmath,verbatim,enumerate}
\usepackage{amssymb}
\usepackage{multirow}
\usepackage{physics}
\usepackage[toc,page]{appendix}
\usepackage{xspace}
\usepackage{slashed}
\allowdisplaybreaks[1]
\newlength{\figurewidth}
\newcommand{\beq}{\begin{equation}}
\newcommand{\eeq}{\end{equation}}
\newcommand{\bea}{\begin{eqnarray}}
\newcommand{\eea}{\end{eqnarray}}
\newcommand{\ba}{\begin{array}}
\newcommand{\ea}{\end{array}}

\newcommand{\pt}{\partial}

\newcommand{\al}{\alpha}
\newcommand{\bt}{\beta}
\newcommand{\g}{\gamma}

\newcommand{\ta}{\theta}
\newcommand{\lam}{\lambda}
\newcommand{\Lam}{\Lambda}
\newcommand{\G}{\Gamma}

\newcommand{\de}{\delta}

\newcommand{\om}{\omega}

\newcommand{\Sg}{\Sigma}

\makeatother
\begin{document}
%
% define Title, Author, Address, Preprint#
%%%%%%%%%%%%%%%%%%%%%%%%%%%%%%%%%%%%%%%%%%%%%%%
\title{Entanglement entropy of non-local theories in AdS}
\setlength{\figurewidth}{\columnwidth}
%%%%%%%%%%%%%%%%%%%%%%%%%%%%%%%%%%%%%%%%%%%%%%%
%
\author{Gaurav Narain$\,{}^a$}
\email{gaunarain@gmail.com}
\author{Nirmalya Kajuri$\,{}^{b}$}
\email{nirmalyak@gmail.com}
\affiliation{
${}^a$ Department of Space Science, Beihang University, Beijing 100191, China.\\
${}^b$ Chennai Mathematical Institute, Siruseri Kelambakkam 603103.
}
%
%
%%%%%%%%%%%%%%%%%%%%%%%%%%%%%%%%%%%%%%%%%%%%%%%
\begin{abstract}
We investigate the effect of non-locality on entanglement entropy 
in anti-de Sitter space-time. We compute entanglement entropy of a 
nonlocal field theory in anti-de Sitter space-time and find several interesting features. 
We find that area law is followed, but sub-leading terms are affected by non-locality. 
We also find that the UV finite term is universal. For the massless theory 
in 3 dimensional AdS we compute it exactly and find the novel feature that it shows oscillatory behavior.
\end{abstract}

\maketitle
%
%%%%%%%%%%%%%%%%%%%%%%%%%%%%%%%%%%%%%%%%%%%%%%%
%

%%%%%%%%%%%%%%%%%%%%%%%%%%%%%%%%%%%%%%%%%%%%%%%
\section{Introduction}
\label{intro}
%%%%%%%%%%%%%%%%%%%%%%%%%%%%%%%%%%%%%%%%%%%%%%%

Entanglement entropy is a very useful tool that helps us quantize the correlation 
between different localized subsystems of a quantum system. 
For local UV-finite theories entanglement entropy is known to follow 
an area law \cite{Bombelli:1986rw, Srednicki:1993im}. Explicit calculation 
in $d > 2$ can be done for free field theories \cite{Casini:2009sr}.
For field theories with holographic duals, entanglement entropy of a 
region $A$ can be computed holographically using the quantum 
corrected Ryu-Takayanagi formula \cite{Ryu:2006ef, Faulkner:2013ana}. 
\beq
\label{qcrt}
S_{bdry}(A) = \frac{\text{Area}_{min}}{4G_N} + S_{bulk}+ \cdots \, .
\eeq
where Area$_{min}$ denotes the minimal area surface in the 
bulk which ends in the boundaries of $A$. The ellipsis refer to higher order 
corrections in $1/G_N$. The last term in the formula is the entanglement 
entropy in AdS. The entanglement entropy of fields in AdS is therefore an 
important component of the quantum corrected Ryu-Takayanagi formula and 
deserves further exploration. So far it has only been calculated for local free 
fields in AdS \cite{Sugishita:2016iel}.

Nonlocal field theories present an interesting situation in this context. One 
would expect that the presence of nonlocal correlations in the theory to 
manifest itself in entanglement entropy. Indeed it is found that explicit 
non-locality can have drastic effects on entanglement entropy including modification 
of area law in some cases \cite{Shiba:2013jja, Karczmarek:2013xxa, Fischler:2013gsa,Pang:2014tpa}, 
but not always \cite{Nesterov:2010yi,Nesterov:2010jh,Solodukhin:2011gn}.
It is therefore interesting to explore the behaviour of
entanglement entropy in field theories which have non-localities.

Nonlocal theories are quite interesting in their own and have been 
explored in a variety of contexts such as 
\cite{Modesto:2011kw,Modesto:2017sdr,Maggiore:2016gpx,Narain:2017twx,Narain:2018hxw}.
Non-local theories have also been explored in the context of black-hole information 
loss paradox \cite{Kajuri:2017jmy,Kajuri:2018myh}.

In this paper we compute entanglement entropy for a particular free non-local field theory.
Our action is given by
\beq
\label{eq:NLact}
S_{NL} = \int {\rm d}x \sqrt{-g}
\biggl[
\frac{1}{2} (\pt \phi)^2 + \frac{m^2}{2} \phi^2 
-\frac{\lam^2}{2} \phi \frac{1}{-\Box} \phi
\biggr] \, ,
\eeq
where we refer to $\lam$ as the non locality scale. It has mass 
dimensions $[mass]^2$. In flat spacetime where it is possible to do a 
momentum decomposition of fields, it should be noticed that 
such non-local terms will become important at small energy.  

 Apart from the general motivations given for computing entanglement 
 entropy in non local field theories, there is a further motivation for this theory. 
 We have shown in a companion paper that these theories admit CFT duals \cite{Kajuri:2018wow}. 
 This calls for further investigation of these theories in a holographic context. 
 Computation of entanglement entropy of this theory in AdS background 
 would be a step towards a full understanding of the quantum corrected 
 Ryu Takayanagi formula \eqref{qcrt} for these theories.  
 
 In this paper we consider odd dimensional anti-de Sitter spaces. The 
 entanglement entropy for the theory \eqref{eq:NLact} for generic odd dimensional 
 AdS is computed using the replica trick and the heat kernel method. There are 
 several interesting features. We find that the area law is followed but the 
 sub-leading terms depend on the non-locality scale. We find that there is a 
 UV-finite universal term in entanglement entropy. For massless theories in 
 AdS$_3$ we explicitly compute it and show that it exhibits oscillatory behaviour. 
 This is a novel feature of nonlocal theories.
 
 The paper is organised as follows. In the section \ref{NLsca} we study the 
 nonlocal theory \eqref{eq:NLact} in more detail. We calculate the Green function 
 for this theory in flat and AdS space-time in section \ref{flat} 
 and \ref{ads}. In the section \ref{EE} we present our computation 
 of entanglement entropy. We summarise our results in the conclusion section \ref{conc}.

%%%%%%%%%%%%%%%%%%%%%%%%%%%%%%%%%%%%%%%%%%%%%%%
\section{Non-local scalar field}
\label{NLsca}
%%%%%%%%%%%%%%%%%%%%%%%%%%%%%%%%%%%%%%%%%%%%%%%

In this section we consider a local scalar field theory consisting of two coupled fields.
This leads to a non-local action once one of the field gets decoupled from the 
system. Consider the following action for local field theory
\beq
\label{eq:slocal}
S = \int {\rm d}x \sqrt{-g} \biggl[
\frac{1}{2} (\pt \phi)^2 + \frac{1}{2} (\pt \chi)^2 + \frac{m^2}{2} \phi^2 
- \lam \phi \chi
\biggr] \, ,
\eeq
where $\phi$ and $\chi$ are two scalar fields on curved non-dynamical 
background, $\lam$ is their coupling strength while $m$ is mass of scalar $\phi$. 
The equation of motion of two fields give $(-\Box + m^2)\phi - \lam \chi=0$ 
and $-\Box \chi - \lam \phi =0$. Integrating out $\chi$ from the second equation 
of motion yields $\chi = \lam (-\Box)^{-1} \phi$. This when plugged back into 
the action (\ref{eq:slocal}) yields a massive non-local theory for scalar $\phi$
whose action is given by eq. (\ref{eq:NLact}).
This non-local action has issues of tachyon. In simple case of 
flat space-time it is noticed that the non-local piece in action reduces 
to $\phi(-p)(-\lam^2/p^2) \phi(p)$ (where $\phi(p)$ is the Fourier 
transform of field $\phi(x)$). This piece correspond to something like 
tachyonic mass thereby resulting in issues of unitarity and 
instability of vacuum. Also, in massless case if we write 
$\phi_1 = (\phi+\chi)/\sqrt{2}$ and $\phi_1 = (\phi-\chi)/\sqrt{2}$, 
then the local action in eq. (\ref{eq:slocal}) can be reduced 
to action of two decoupled free scalar scalar fields 
$\phi_1$ and $\phi_2$ of mass $\lam$ and $-\lam$ respectively. 
Immediately we notice the presence of tachyon field $\phi_2$. 
In the massive case things are just more involved, but tachyonic 
issue remains. In what follows we will therefore consider
$\lam^2 \to -\lam^2$. This overcomes the issue of tachyons 
in the non-local theory. Also, it should be mentioned that 
the resulting non-local theory with $-\lam^2$ can also arise as a low-energy 
limit of some ultraviolet complete theory in which sense 
it is an effective theory. 

In this section we will study this simple free non-local theory 
on flat and AdS space-time. In generic space-time the 
green's function equation is given by,
\beq
\label{eq:NLgreenFeq}
\biggl(
-\Box + m^2 + \frac{\lam^2}{-\Box}
\biggr) G(x,x^\prime) = -\frac{i \de(x-x^\prime)}{\sqrt{-g}} \, .
\eeq

%%%%%%%%%%%%%%%%%%%%%%%%%%%%%%%%%%%%%%%%%%%%%%%
\subsection{Flat space-time}
\label{flat}
%%%%%%%%%%%%%%%%%%%%%%%%%%%%%%%%%%%%%%%%%%%%%%%

In this section we study this theory on flat space-time. Understanding the 
behaviour of theories in flat space-time is important as in curved space-time 
at ultra-local distances the theory behaves as in flat. This is also 
required for implementing suitable boundary conditions for 
the computation of the green's function. 

In flat space-time one can use the momentum space 
representation to write the propagator. Writing 
$G(x,x^\prime) = \int {\rm d}p/(2\pi) \tilde{G}(p) 
\exp\{ip(x-x^\prime)\}$, it is seen that 
\beq
\label{eq:GflatNL}
G^{\rm NL}_f(x,x^\prime)
= -i \int \frac{{\rm d}^dp}{(2\pi)^d}
\biggl(
p^2 + m^2 + \frac{\lam^2}{p^2}
\biggr)^{-1} e^{i p(x-x^\prime)}
\eeq
where the subscript in $G_f$ implies flat space-time. The integrand 
can be integrated easily by making use of partial fraction 
and decomposing it. It can be written in an alternative manner as:
\beq
\label{eq:partDe}
\frac{1}{p^2 + m^2 + \lam^2/p^2}
= \frac{p^2}{p^4 + m^2 p^2 + \lam^2}
=\frac{A}{p^2 + r_{-}^2} + \frac{B}{p^2+r_{+}^2} \, ,
\eeq
where 
\beq
\label{eq:RpRm}
r_{-}^2 = \frac{m^2 - \sqrt{m^4 - 4\lam^2}}{2} \,
\hspace{5mm} 
r_{+}^2= \frac{m^2 + \sqrt{m^4 - 4\lam^2}}{2} \, ,
\eeq
while 
$A = - r_{-}^2/(r_{+}^2 - r_{-}^2)$ and $B= r_{+}^2/(r_{+}^2 - r_{-}^2)$. 
For massless fields ($m^2\to0$) $r_{\pm}^2=\pm i \lam$, $A=1/2$ 
and $B=1/2$. After the partial fraction each of the 
piece can be further expressed by using inverse 
Laplace transform. This allows us to write 
$\bigl(p^2 + m^2 + \lam^2/p^2\bigr)^{-1}$ in inverse Laplace form. 
Then one can express the flat space-time 
green's function in eq. (\ref{eq:GflatNL}) as follows 
\bea
\label{eq:invLapflat}
G^{\rm NL}_f(x,x^\prime)
= -i \int \frac{{\rm d}^dp}{(2\pi)^d}
\int_0^{\infty} {\rm d}s 
\bigl[
\cosh(s \bt) - \frac{\al}{\bt} \sinh(s \bt)
\bigr] 
e^{-s(p^2+\al)+i p(x-x^\prime)} \, ,
\eea
where $\al=m^2/2$ and $\bt=\sqrt{m^4-4\lam^2}/2$. 
In the integrand one can smoothly take massless limit to 
obtain green's function for non-local 
massless scalar-field. In this massless limit $\al \to0$
while $\bt\to i\lam$, then the term in square bracket in the 
integrand still remains real as $\cosh(s\bt) \to \cos(s\lam)$
and $\sinh(s\bt) \to i \sin(s\lam)$. 

In Eq. (\ref{eq:invLapflat}) one can first evaluate the $p$-integration 
which is done by completing the square leading to a gaussian type 
integral. In the process of completing square for $p$, one encounters 
division by $s$. This implies that the lower limit of $s$-integration should be 
regularised by imposing a cutoff. One can then perform 
integration over $p$ easily leading to,
\beq
\label{eq:pintDone}
G^{\rm NL}_f(x,x^\prime) = \int_{\de^2}^{\infty} {\rm d}s 
(4\pi s)^{-d/2} e^{- \mu^2/4s} \left(
A e^{-s r_{-}^2} + B e^{-s r_{+}^2}
\right) \, ,
\eeq
where $\mu(x,x^\prime) = \sqrt{(x-x^\prime)^2}$. 
This is a very subtle definite integral due to the lower limit of integration.
This integral can be seen as a sum of two integrals of 
local theories with masses $r_{-}$ and $r_{+}$ respectively. 
It is important to understand the singularity structure of 
this integral carefully as it will be seen to a play 
an important role later on. As a result we first consider the local 
field theory which can be obtained from above by taking 
$\lam\to0$ limit. The above integral becomes following 
\beq
\label{eq:GfLocal}
G^{\rm L}_f(x,x^\prime) = \int_{\de^2}^{\infty} {\rm d}s 
(4\pi s)^{-d/2} e^{- \mu^2/4s} e^{-s m^2} \, .
\eeq
When $x\neq x^\prime$ ($\mu\neq0$), the integral can be 
performed exactly in closed form for $\de=0$. Even otherwise, it is 
seen that $G^{\rm L}_f$ has a taylor series expansion around 
$\de=0$. The leading term of which has a singularity 
which goes as $\sim \mu^{2-d}$. On the other hand it is seen 
that performing first the small distance expansion and then 
doing the $s$-integration leads to no short distance singularity in $G_f$. 
This indicates the subtlety associated with the point $(\de,\mu) = (0,0)$.
The standard rule we will follow is to first perform the $s$-integration 
and then extract the short-distance singularity. 

Keeping this in mind eq. (\ref{eq:invLapflat}) is evaluated by first performing
the integration over $p$ then integration over $s$. 
For $m=0$ case the $s$-integral 
can be performed exactly in closed form in arbitrary dimensions. 
In four dimensions it acquires a simplified form and is given by,
\beq
\label{eq:gflatNL}
G_f(x,x^\prime) =
\frac{\sqrt{i \lam}K_1\left(\sqrt{i \lam}\mu \right)+\sqrt{-i \lambda } 
K_1\left(\sqrt{-i \lam} \mu \right)}{8 \pi ^2 \mu } \, .
\eeq
In the case when $m\neq0$ things are complicated. Here two possibilities arises 
$m^4>4\lam^2$ and $4\lam^2>m^4$, where both $m$ and $\lam$ are positive. 
In the former case one can take limit $\lam\to0$ (locality limit), while in later case 
one can take the limit $m\to0$ (massless limit). In the locality limit, $r_{-}^2=0$
and $r_{+}^2=m^2$. In this case $A=0$ while $B=1$ ($\al=\bt=m^2/2$). In this case 
one gets the propagator for local massive scalar field which is  
\beq
\label{eq:gfmlam0}
\left. G^{\rm NL}_f (\mu)\right|_{\lam=0} = 
(2 \pi )^{-d/2} \left(\frac{m}{\mu}\right)^{(d-2)/2}
K_{\frac{d}{2}-1}(m \mu) = G_f^L(\mu)\, ,
\eeq
where $G_f^L(\mu)$ is the local green's function of free massive scalar field.
For $m\neq0$ one can still perform integral in closed form. 
For the massive non-local scalar field in arbitrary space-time dimension 
this green's function $G_f$ is given by
\bea
\label{eq:Gfmlam}
G^{\rm NL}_f(\mu)=
\frac{(2\pi)^{-d/2}\mu^{1-d/2}}{(r_{+}^2-r_{-}^2)} \times
\biggl[
r_{+}^{d/2+1} K_{\frac{d}{2}-1} \left(r_{+}\mu\right)
-r_{-}^{d/2+1} K_{\frac{d}{2}-1} \left(r_{-}\mu\right)
\biggr] \, ,
\eea
where $r_{+}$ and $r_{-}$ are stated before. In fact one can express this 
green's function as a sum of green's function for two local massive scalar fields with mass 
$r_+^2$ and $r_{-}^2$ and, coefficient $A$ and $B$ respectively. This is 
expected and is clear from eq. (\ref{eq:pintDone}). As a result one can also 
write $G_f^{\rm NL}$ as
\beq
\label{eq:GfsumScalar}
G^{\rm NL}_f(\mu)= \left. A G_f^L (\mu) \right|_{r_{-}}
+ \left. B G_f^L (\mu) \right|_{r_{+}} \, .
\eeq
This will also be useful in fixing the appropriate boundary 
conditions for the green's function on AdS. 
One can do a short distance expansion of the flat space-time 
green's function to isolate the singular behaviour of the 
propagator. In four dimensions the two singular contributions are:
\beq
\label{eq:Gfsing}
G_f^{\rm sing} = 
\frac{1}{4\pi^2 \mu^2} + \frac{m^2}{8\pi^2} \log(\mu) + \cdots \, ,
\eeq
where $\mu=\mu(x,x^\prime)$. These are singular behaviour in four dimension
at short distances. This will be true in other even dimensions where the leading 
singularity will be $\sim \mu^{2-d}$. In even and odd dimensions we have the 
short distance singularity structure as
\begin{align}
\label{eq:GfsingEvenOdd}
& \left. G_f^{\rm sing} \right|_{\rm d=even} 
= a_0 \mu^{2-d} + a_1 \mu^{4-d} + \cdots a_L \log \mu \, ,
\notag \\
& \left. G_f^{\rm sing} \right|_{\rm d=odd} 
= a_0 \mu^{2-d} + a_1 \mu^{4-d} + \cdots + \frac{a_L}{\mu} \, .
\end{align}
The $\log \mu$ contribution in even dimensions is also singular though 
in infrared limit ($\mu\to\infty$) it has a well-defined behaviour. 
In massive theories one also has 
an inbuilt infrared cutoff $\Lam_{\rm IR} \sim 1/m$, which 
truncates the IR modes of the theory. Then it is noticed 
that the $\log \mu$ approaches $\log \Lam_{\rm IR} = - \log m$ in IR.
This implies that the $\log$ contribution in $G_f \sim -m^2 \log m$ which 
has a smooth massless limit. 

%%%%%%%%%%%%%%%%%%%%%%%%%%%%%%%%%%%%%%%%%%%%%%%
\subsection{Anti-DeSitter (AdS)}
\label{ads}
%%%%%%%%%%%%%%%%%%%%%%%%%%%%%%%%%%%%%%%%%%%%%%%

In this section we will investigate the theory eq. (\ref{eq:NLact}) in Euclidean AdS (EAdS) background.
The $n$-dimensional EAdS space-time $\mathbb{H}^n$ can be identified with the real 
upper-sheeted hyperboloid in $(d+1)$ Minkowski space-time $M_{d+1}$:
$X_d= \{x \in \mathbb{R}^{d+1}, x_1^2+\cdots +x_d^2-x_{d+1}^2= - H^{-2}\}$, where $H$ is 
related to EAdS length. 
If the two points are denoted by $x$ and $x^\prime$, the length of 
geodesic connecting them is $\mu(x,x^\prime)$. It is useful to introduce 
a quantity $z(x,x^\prime)=\cosh^2(H\mu/2)$. For time-like distances 
$\mu^2<0$ which corresponds to $0\leq z<1$, 
while space-like separation $0<\mu^2<\infty$ corresponds to $1<z<\infty$. 
$z=1$ is the light-like separation.
On AdS background the Green's function will be entirely 
a function of $\mu$ (or $z(x,x^\prime)$). 
This allows one to compute green's function 
exactly by solving linear differential equations. 

In the case of non-local theory, the green's function equation is given in 
eq. (\ref{eq:NLgreenFeq}). Before we solve for $G(x,x^\prime)$
on AdS, we first note the following important identities on curved 
space-time. 
\beq
\label{eq:dsOPiden}
-[\Box^2 - m^2 \Box + \lam^2]^{-1}\Box
= A (-\Box + r_{-}^2)^{-1} + B (-\Box + r_{+}^2)^{-1} \, ,
\eeq
where $A$, $B$, $r_{-}^2$ and $r_{+}^2$ have the same values 
as in flat space-time.
If we multiply both sides by operator 
\beq
\label{eq:opProd}
[\Box^2 - m^2 \Box + \lam^2]
= (-\Box + r_{-}^2)(-\Box + r_{+}^2) \, ,
\eeq
then we clearly note that 
it is identically satisfied. It is this identify which allows us to compute the 
green's function of the non-local theory in anti-deSitter space-time. 
This means that the full green's function for non-local theory 
$G_{\rm NL}(x,x^\prime)$ is a sum of two green's function 
\beq
\label{eq:GNLsum}
G_{\rm NL}(x,x^\prime) = A G_{-}(x,x^\prime) + B G_{+}(x,x^\prime) \, ,
\eeq
where 
\beq
\label{eq:G1G2eq}
(-\Box + r_{-}^2)G_{-}(x,x^\prime)=0 \, , 
\hspace{5mm}
(-\Box + r_{+}^2)G_{+}(x,x^\prime)=0 \, 
\eeq
are the two green's function on AdS which are determined 
by requiring that the short distance singularity of 
AdS propagator to match with the singular behaviour of the 
flat space-time propagator. This is just gives the usual 
AdS propagator of the free local massive scalar field 
which is well known in literature \cite{Avis:1977yn,Allen:1985wd,Burgess:1984ti,Caldarelli:1998wk}.
For example, for local massive scalar field on AdS we have 
$(-\Box+m^2)G(x,x^\prime)=0$ as Green's function equation. 
Then $G(x,x^\prime)$ follows 
\beq
\label{eq:GmOP}
z(1-z)G^{\prime\prime}
+\left[c-(a+b+1)z\right] G^\prime - ab G=0 \, ,
\eeq
where $a=\bigl[d-1+\bigl((d-1)^2 + 4m^2/H^2 \bigr)^{1/2}\bigr]/2$, 
$b=\bigl[d-1-\bigl((d-1)^2 + 4m^2/H^2 \bigr)^{1/2}\bigr]/2$
and $c=d/2$. This Hyper-geometric differential equation 
has two linearly independent solution:
$z^{-a}{}_2F_1(a,a-c+1,a-b+1;z^{-1})$ and $z^{-b}{}_2F_1(b,b-c+1,b-a+1;z^{-1})$ 
\cite{Avis:1977yn,Allen:1985wd,Burgess:1984ti,Caldarelli:1998wk,Gubser:2002zh}. 
As $a>0$ and $b<0$, therefore the former solution 
fall off at boundary of AdS while the later doesn't. 
This allow one to pick the former.
By requiring the short distance singularity of 
AdS propagator to match with the singular behaviour of the 
flat space-time propagator one concludes 
that $G(z) = q \times z^{-a}{}_2F_1(a,a-c+1,a-b+1;z^{-1})$.
The coefficient $q$ is fixed by requiring that 
the strength of singularity of AdS propagator matches with the strength of 
singularity in flat space-time. This gives,
\beq
\label{eq:Qads}
q = \frac{(4\pi)^{(1-d)/2} 2^{-(a-b+1)} H^{d-2} \G(a)}{\G(a-c+3/2)} \, .
\eeq 
Using this one can write the Green's function $G_{NL}$ for 
non-local scalar on AdS to be
\bea
\label{eq:GNLmdSform}
&&
G_{\rm NL}(z) = A q_{-} \, z^{-a_{-}}{}_2F_1(a_{-},a_{-}-c_{-}+1,a_{-}-b_{-}+1;z^{-1})
\notag \\
&&
+ B q_{+} \,  z^{-a_{+}}{}_2F_1(a_{+},a_{+}-c_{+}+1,a_{+}-b_{+}+1;z^{-1})\, ,
\eea
where the coefficients $q_{-}$ and $q_{+}$ are determined from 
eq. (\ref{eq:Qads}) with mass replaced by $r_{-}$ and $r_{+}$ 
respectively, while $A$ and $B$ are same 
as previous mentioned. The parameters of the hyper-geometric 
functions are given by
\bea
\label{eq:abc}
&&
a_{-} = \frac{1}{2} \left(d-1+\sqrt{(d-1)^2 + \frac{4 r_{-}^2}{H^2} } \right) \, ,
\hspace{2mm}
b_{-} = \frac{1}{2} \left(d-1-\sqrt{(d-1)^2 + \frac{4 r_{-}^2}{H^2} } \right) \, ,
\notag \\
&&
a_{+} = \frac{1}{2} \left(d-1+\sqrt{(d-1)^2 + \frac{4 r_{+}^2}{H^2} } \right) \, ,
\hspace{2mm}
b_{+} = \frac{1}{2}  \left(d-1-\sqrt{(d-1)^2 + \frac{4 r_{+}^2}{H^2} } \right) \, ,
\eea
and $c_{-} = c_{+} = d/2$. The parameters $q_{-}$ and $q_{+}$ are given by,
\beq
\label{eq:Q1Q2}
q_{-} = \frac{2^{-(a_{-}-b_{-}+1)} H^{d-2} \G(a_{-})}{(4\pi)^{(d-1)/2}\G(a_{-}-c_{-}+3/2)} \, ,
\hspace{3mm}
q_{+} = \frac{2^{-(a_{+}-b_{+}+1)} H^{d-2} \G(a_{+})}{(4\pi)^{(d-1)/2}\G(a_{+}-c_{+}+3/2)} \, .
\eeq
The green's function is analytic in $\lam$. By taking $\lam \to i\lam$ one 
obtains green's function for theory with tachyons. The green's function 
remains well-defined even when this transformation is done. 
The massless limit can be smoothly taken on AdS irrespective of non-locality. 
In the massless limit $r_{+}^2 \to i \lam$ while $r_{-}^2\to -i\lam$,
while $A\to1/2$ and $B\to1/2$. Correspondingly the parameters 
of hyper-geometric functions can be obtained
(see \cite{Narain:2018rif} for non-local green's function 
in deSitter). 

\begin{figure}[h]
\centerline{
\vspace{0pt}
\centering
\includegraphics[width=4.3in,height=3in]{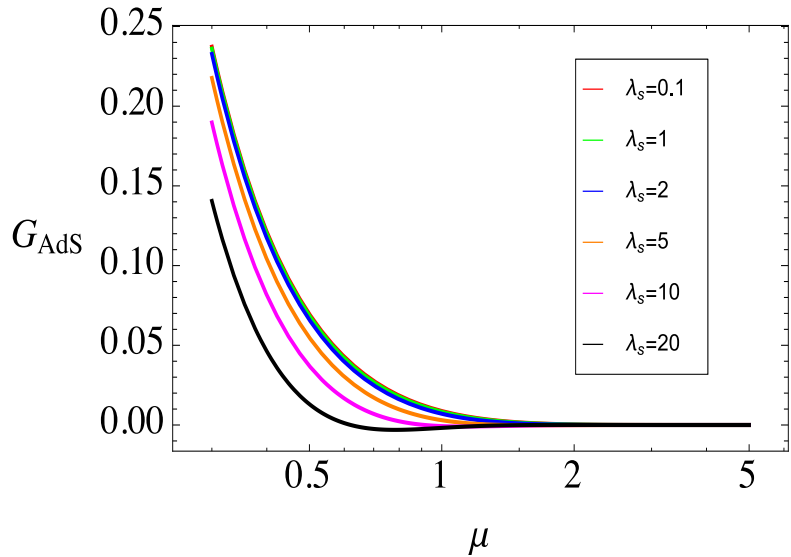}
}
\vspace{-3mm}
\caption[]{
The Green's function of massless non-local scalar field in four space-time 
dimensions. The propagator is plotted for various strength of non-locality $\lam$
against the for space-like separation $\mu(x,x^\prime)$. 
}
\label{fig:Gm0adsNL}
\end{figure}

\begin{figure}[h]
\centerline{
\vspace{0pt}
\centering
\includegraphics[width=4.3in,height=3in]{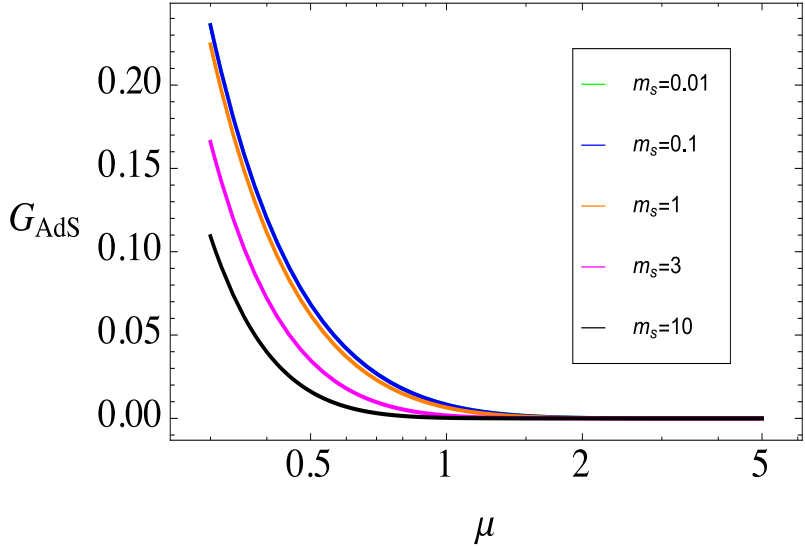}
}
\vspace{-3mm}
\caption[]{
The Green's function of massive non-local scalar field in four space-time 
dimensions. The propagator is plotted for various mass $m_s$
and for fixed $\lam_s=1$, 
against the for space-like separation $\mu(x,x^\prime)$. 
}
\label{fig:GmNL_varm_1mu}
\end{figure}

In Figure \ref{fig:Gm0adsNL} we plot this green's function 
for various values of $\lam_s$ for massless theory. It is seen 
that non-locality starts to have effect on the propagator at 
short distances. 
%
%%%%%%%%%%%%%%%%%%%%%%%%%%%%%%%%%%%%%%%%%%%%%%%%%%%%%%%
%%%%%%%%%%%%%%%%%%%%%%%%%%%%%%%%%%%%%%%%%%%%%%%%%%%%%%%
%
For the case of non-zero mass, the propagator has one more parameter, however 
structure remains same. Here we have three cases
(as in flat space-time): $m^2<2\lam$, $m^2=2\lam$ and $m^2>2\lam$. 
In each case the propagator is real for space-like separation.
It is worthwhile to plot the propagator for fixed value of $\lam_s$ and 
for decreasing mass. It is seen that as $m\to0$, the massive 
propagator smoothly approaches the massless non-local propagator. 
In figure \ref{fig:GmNL_varm_1mu} we plot this scenario. 
%
%%%%%%%%%%%%%%%%%%%%%%%%%%%%%%%%%%%%%%%%%%%%%%%%%%%%%%%
%%%%%%%%%%%%%%%%%%%%%%%%%%%%%%%%%%%%%%%%%%%%%%%%%%%%%%%
%
The case of equality is interesting as then the propagator depends only on 
one parameter. Physically it implies that the two length scales are 
comparable. In this case the non-locality scale is coupled with the mass, so 
$m\to0$ implies $\lam\to0$ simultaneously. The infrared limit 
of mass going to zero is well-defined and smooth. In this limit the 
non-local green's function is independent of any parameters in the theory
(depending only on $H$). It is given by,
\beq
\label{eq:GNL_lamMeq}
\left. G_{\rm NL} \right|_{m^2 = 2\lam} = 
\frac{H^2 \left[2 z+2 (z-1) z \log \left(\frac{z-1}{z}\right)-1\right]}{16 \pi ^2 (z-1) z} \, .
\eeq
On doing the short distance expansion $\mu\to0$ it is noticed that 
the leading singular behaviour of $G_{\rm AdS}(\mu)$ is exactly the same as in
flat space-time, which is expected due to boundary conditions. 
Beside this there is next-to-leading singular piece in four 
space-time dimensions. This term is $\log$-divergent and 
is given by $((2H^2 + m^2)/8\pi^2) \log \mu$. Compared to 
flat space-time, here there is extra contribution from $H$. 
Again, this term doesn't depend on the non-locality strength $\lam$. 
This again is not problematic in IR as argued in flat space-time. 
Beside this there will also be terms like $\mu \log\mu$ which will appear 
in higher-order expansion of the green's function. Terms like these are 
not UV-singular in the short distance limit. It is noted that in short-distance 
expansion one can arrange the various terms as the following series 
\beq
\label{eq:G_NLexp}
G_{\rm NL} = - \frac{\pi^{1-d/2} 2^{-d} \csc(d\pi/2)}{
H^{d-2} \G(2-d/2)} \frac{1}{(z-1)^{d/2-1}} + \frac{E_2}{(z-1)^{d/2-2}} + \cdots + E_0 \, ,
\eeq
where $E_0$ is a constant piece in $z$ depending on mass $m$, 
non-locality strength $\lam$, $H$ and dimension $d$. This is a kind of 
universal piece as it is present in the expression of $G_{\rm NL}$ for all $z$. 
This term is given by,
\beq
\label{eq:E0}
E_0 = \frac{H^{d-2} \pi^{1-d/2} \csc(d\pi/2)}{2^{d} \G(d/2) (r_{+}^2 - r_{-}^2)} 
\left[\frac{r_{+}^2 \G(a_{+})}{\G(1-b_{+})}
-  \frac{r_{-}^2 \G(a_{-})}{\G(1-b_{-})}
\right] \, .
\eeq
%

%%%%%%%%%%%%%%%%%%%%%%%%%%%%%%%%%%%%%%%%%%%%%%%
\section{Entanglement entropy}
\label{EE}
%%%%%%%%%%%%%%%%%%%%%%%%%%%%%%%%%%%%%%%%%%%%%%%

In this section we compute the entanglement entropy (EE) in the AdS background 
using the replica trick \cite{Callan:1994py,Holzhey:1994we,Calabrese:2004eu,Calabrese:2009qy}. 
Replica trick is covariant formulation of the computation 
of either EE or Renyi entropy via usage of effective action of the theory
which is computed on a background with conical defect. 

\begin{figure}
\label{fig:Bound}
\centering
\begin{tikzpicture}
\draw[fill=lightgray] (0,2) arc (270:90:-2);
\node[below left] at (0,0) {$B$};
\draw[-latex,thick,black] (0,0)--(4,0);
\node[below ] at (4,0) {$z$};
\draw[-latex,thick,black] (0,-3)--(0,3);
\node [above left] at (0,3) {$x$};
\node [below right] at (0.5,1.2) {$A$};
\node [above right] at (1,1.8) {$\Sigma$};
\node [below right] at (2.0,1.2) {$\bar{A}$};
\end{tikzpicture}
\caption{Minimal surface corresponding to region $B$ in the boundary. The $z$-direction 
denotes the bulk while the $x$ represent the spatial co-ordinate on boundary. The region 
surrounded by $B$ is $\Sg$ and is denoted by $A$.}
\end{figure}
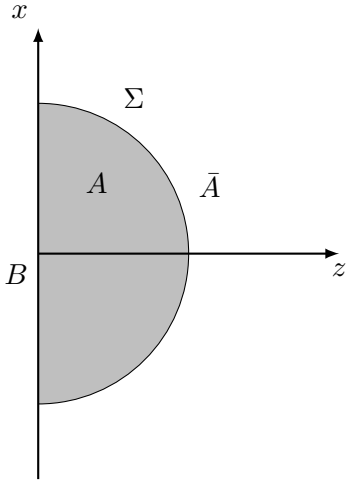

The actual starting point is density matrix $\rho$. Any space can be divided 
by a surface $\Sigma$ in to two parts: $A$ and $\bar{A}$. The field modes resides on 
both the sides of the boundary surface $\Sigma$ (the projection of this 
on boundary is $B$). This has been depicted in the figure \ref{fig:Bound}. 
This leads to a Hilbert 
space factorisation: $H = H_A \otimes H_{\bar{A}}$. Then tracing out degree 
of freedom on $\bar{A}$ result in reduced density matrix:
$\rho_{A} = {\rm Tr} [\ket{\psi} \bra{\psi}]$, where $\bar{\psi}$ is a 
ground state. The entanglement entropy $S_{A}$ will be defined 
as: $S_{A} = - {\rm Tr} (\rho_A \log \rho_A)$. In order to compute 
EE for field theoretic systems one has to make use of 
replica trick, which make use of path-integral formalism. 
In this one start by expressing the density matrix 
in the path-integral form. Following \cite{Ryu:2006ef} 
we first define a vacuum state of QFT by path-integral over 
half of total Euclidean space $\tau\leq0$, in such a manner that the
quantum field takes a fixed boundary condition 
$\psi(\tau=0,x)=\psi_0(x)$. Then we have
\beq
\label{eq:psiBd}
\Psi[\psi_0(x)] = \int_{\psi(\tau=0,x)=\psi_0(x)} {\cal D} \psi \exp(-I) \, ,
\eeq
where $I$ is the euclidean action of the field $\psi$ and 
$\Psi$ is a functional of boundary $\psi_0$. The direction of 
$x$ is orthogonal to boundary $\Sigma$. The surface defined by 
$x=0$ , $\tau=0$ naturally divides the surface $\Sigma$ in two parts:
$x>0$ and $x<0$. Correspondingly the boundary data gets separated 
as $\psi_{\pm}(x) = \psi_0(x)$ for $x>0$ and $x<0$ respectively. Tracing out 
over $\psi_{-}$ results in a reduced density matrix
\beq
\label{eq:redDenrho}
\rho(\psi_{+}^1, \psi_{+}^2)
= \int {\cal D} \psi_{-} \Psi(\psi_{+}^1, \psi_{-})
\Psi(\psi_{+}^2, \psi_{-}) \, ,
\eeq
where the path-integral goes over whole euclidean space-time 
except the cut at $\tau=0$, $x>0$, while the field $\psi(A)$ takes 
the value $\psi^2_{+}$ above the cut and $\psi^1_{+}$ below the cut. 
The quantity ${\rm Tr} \rho^n$ is then defined over $n$-sheeted 
copy of the cut space-time, where appropriate analytic gluing is done 
when passing from one sheet to another. This quantity is given by,
\beq
\label{eq:trrhoN}
{\rm Tr} \rho^n = \frac{Z_n}{Z_1^n} \, ,
\eeq
where $Z_n$ is the path-integral over the $n$-sheeted covering space
${\cal M}^n$ which is obtained by analytically sewing $n$-copies of 
the original euclidean manifold ${\cal M}$ along the cut. The 
EE and Renyi entropy are correspondingly given by 
\beq
\label{eq:EERen}
S_{EE} = \lim_{n\to1} (n \pt_n - 1) W_n \, ,
\hspace{5mm}
S_n = \lim_{n\to1} \frac{W_n - n W_1}{n-1}  \, ,
\eeq
where $W_n = -\log Z_n$ is the effective action on ${\cal M}^n$
while $W_1 = \log Z_1$ is the usual effective action on ${\cal M}$. 

The effective action $W_1$ (where $1$ indicates the EA has been computed on the 
single-copy of manifold ${\cal M}$) for field theory can be computed 
using the standard field theory methods in flat space-time. On curved 
space there are limitations as the usual Feynman perturbation theory 
cannot be directly extended. However, one can still perform some 
computations on maximally symmetric background or one-loop approximation. 
In the later case, the effective action $W_1$ to one-loop can be expressed 
using the heat-kernel of the Hessian of the theory on the curved background. 
\beq
\label{eq:1loopW}
W = - \frac{1}{2} \int_{0}^\infty \frac{{\rm d}s}{s} {\rm Tr} K(x,x;s)
= -\lim_{x\to x^\prime} \frac{1}{2} \int_{0}^\infty \frac{{\rm d}s}{s} {\rm Tr} 
\bra{x}\exp(-s\triangle)\ket{x^\prime} \, ,
\eeq
where $K(x,x^\prime;s)$ is the heat-kernel of the operator $\triangle$, 
while $x$ and $x^\prime$ are two space-time points and ${\rm Tr}$ is 
over all Lorentz and space-time index. 
The lower limit of the $s$-integral is not well-defined and leads to 
UV divergences which also arise due to co-incident limit.
In fact the integrand is singular at $s=0$. One has to impose a cutoff 
which correspond to imposing an ultraviolet cutoff in theory. 
This singles out precisely the divergent 
part of the above integral leaving behind a UV-finite piece. 

In the replica-method, the entropy is computed by first obtaining the 
effective action on the background of $n$-manifold (${\cal M}^n$). 
This can be computed by exploiting the 
known results of heat-kernel of operator on background of 
cone \cite{Dowker:1977zj,Dowker:1987mn,Solodukhin:2011gn,Fursaev:1994in} 
and making use of Sommerfeld formula \cite{Sommerfeld:1897} which expresses 
heat-kernel modification on the such a manifold with conical singularity as
\bea
\label{eq:HKmodi}
K_n(\phi,\phi^\prime;s) = K_1(\phi,\phi^\prime;s)
+ \frac{i}{4\pi n} \int_{\G} {\rm d} \om \cot\left(\frac{\om}{2n}\right)
K_1 (\phi-\phi^\prime+\om;s) \, ,
\eea
where $\phi$ is the angle and other co-ordinates have been skipped,
the contour $\G$ consists of two lines: one goes from $(-\pi +i \infty)$ to $(-\pi -i \infty)$ 
intersecting the real axis between the poles $-2 \pi n$ and 0 of $\cot(\om/2 n)$,  
and another goes from $(\pi -i \infty)$ to $(\pi +i \infty)$ 
intersecting the real axis between the poles 0 and $2 \pi n$. 
For $n=1$ the integrand is $2\pi$-periodic 
and the contribution from the two vertical lines cancel each other.

This modification of heat-kernel on ${\cal M}^n$ gets reflected on the 
computation of one-loop effective action and the green's function 
on ${\cal M}^n$, and allows to do the computation of the entropy 
via replica trick. In the case of AdS one then has to go to co-ordinate 
system where it is possible to successfully implement the procedure 
of having $n$-copies of manifold where gluing is done analytically to 
have construct ${\cal M}^n$. The maximally symmetric EAdS 
can be represented as an embedding of surface 
\beq
\label{eq:embed1}
x_1^2 + \cdots + x_d^2 - x_{d+1}^2 = -H^{-2} 
\eeq
in 
$\mathbb{R}^{1,d}$ with metric 
\beq
\label{eq:adsMet}
{\rm d}s^2 = {\rm d}x_1^2 + \cdots + {\rm d}x_d^2 - {\rm d}x_{d+1}^2 \, .
\eeq
Under the transformation $x_{d+1} + x_d = z^{-1} H^{-2}$, $x_a = (Hz)^{-1} y_a$, the 
induced metric on the above hyperboloid then becomes a standard 
$AdS_d$ metric: 
\beq
\label{eq:stanAds}
{\rm d}s^2 = (Hz)^{-2} ({\rm d}z^2 + \eta_{ab} {\rm d}y^a {\rm d} y^b) \, .
\eeq
If one polar transforms $z$ and $y$, implying $z=r \sin \ta$ and $y=r \cos \ta$, then 
the entangling surface $\Sg$ is given by $r=r_0$ and $t_E=0$. 
$EAdS_{d}$ has another 
useful foliation given by: 
\bea
\label{eq:useAdS}
&& x_{d+1} = \sqrt{\rho^2 +L^2} \cosh u \, , 
\hspace{2mm}
x_1 = \rho \sin(\tau/L),  
\hspace{2mm}
x_d= \rho \cosh(\tau/L) \, , 
\notag \\
&&
x_2 = \sqrt{\rho^2 +L^2} \sinh(u) \cos(\phi_1) \, ,
\hspace{2mm}
x_3 = \sqrt{\rho^2 +L^2} \sinh(u) \sin(\phi_1) \cos(\phi_2) \, , 
\hspace{2mm}
\cdots, 
\notag \\
&&
x_{d-2} = \sqrt{\rho^2 +L^2} \sinh(u) \sin(\phi_1)
\sin(\phi_2) \cdots \sin(\phi_{d-2}) \, ,
\eea
where $L=H^{-1}$ \cite{Casini:2011kv,Hung:2014npa}. This transformation 
expresses the EAdS space-time as a Euclidean topological black-hole
whose horizon is hyperbolic space $H^{d-2}$. This metric is given by,
\bea
\label{eq:adstopobh}
{\rm d}s^2 =
\frac{{\rm d}\rho^2}{1+H^2 \rho^2}+H^2 \rho^2 {\rm d}\tau_E^2
+(H^{-2}+\rho^2)({\rm d}v^2 + \sinh^2 v \, {\rm d}\Omega_{d-3}^2)\,. 
\eea
where the horizon occurs at $\rho=0$ having a uniform negative curvature. 
This topological black hole has a horizon temperature $T = H/(2\pi)$.
The entangling surface $\Sg$ correspond to this horizon at 
$\tau_E=0$. We represent the area of this entangling surface by 
${\cal A}(\Sg)$ which has been computed in \cite{Casini:2011kv,Sugishita:2016iel}. 
For the $n$-copies of this manifold, the covering space ${\cal M}^n$ 
has the period $\tau_E \sim \tau_E + 2n\pi/H$. The partition function 
$Z_n$ acquires a thermal nature with temperature $H/(2n\pi)$. 

On the AdS geometry, which is maximally symmetric, one can 
compute the heat-kernel $K_1(x,x^\prime;s)$ exactly \cite{Camporesi:1990wm}, 
using which one can compute the effective action by making use 
of the relation eq. (\ref{eq:1loopW}). This can then be used to compute the entropy. 
The effective action involves trace of the heat-kernel. Using the above 
co-ordinate system eq. (\ref{eq:adstopobh}) the trace of the heat-kernel is 
given by,
\begin{align}
\label{eq:traceHKn}
&
{\rm Tr} K_n(s) = H^{3-d} \int_0^{2\pi n/H} {\rm d}\tau_E 
\int_{H^{d-2}} {\rm d}V_{d-2} \int_0^\infty {\rm d}\rho \, \rho(1 + H^2 \rho^2)^{\frac{d-3}{2}}
K_n(x,x;s) \notag \\
& = n {\rm Tr} K_1(s) + \frac{i}{2} \frac{{\cal A}(\Sg)}{H^{d-2}} 
\int_0^\infty {\rm d}\rho \rho (1+ H^2 \rho^2)^{\frac{d-3}{2}}
\int_{\G} {\rm d} \om \cot\left(\frac{\om}{2n}\right) K_1(\mu;s) \, ,
\end{align}
where 
\beq
\label{eq:zEEdef}
z(x,x^\prime) = \cosh^2\left(\frac{H\mu}{2}\right) 
= 1 + (H\rho)^2 \sin^2(H(\tau_E - \tau_E^\prime))
= 1 + (H\rho)^2 \sin^2(\om/2)
\eeq
is the EAdS invariant quantity for the points two $x$ and $x^\prime$ which differ
only in $\tau_E$-direction for the EAdS metric stated in eq. (\ref{eq:adstopobh}), 
while $\mu(x,x^\prime)$ is the geodesic distance between $x$ and $x^\prime$. 
The expression for the trace of heat-kernel immediately give us the expression 
for the effective action on ${\cal M}^n$ using eq. (\ref{eq:1loopW}) which can be 
used to compute entropy. The expression of the effective action is given by,
\beq
\label{eq:Wnexp}
W_n = n W_1 - \frac{i}{2} \frac{{\cal A}(\Sg)}{H^{d-2}} 
\int_{\de^2}^\infty \frac{{\rm d} s}{s} 
\int_0^\infty {\rm d}\rho \rho (1+ H^2 \rho^2)^{\frac{d-3}{2}}
\int_{\G} {\rm d} \om \cot\left(\frac{\om}{2n}\right) K_1(\mu;s) \, .
\eeq
The EE (or Renyi) can be computed following eq. (\ref{eq:EERen}) which is then
given by,
\begin{align}
\label{eq:EEexp}
&
S_{\rm EE} = \lim_{n\to1}
-\frac{-i {\cal A}(\Sg)}{2 H^{d-2}}  
(n\pt_n-1) F(n) \, , 
\\
\label{eq:Renyiform}
& 
S_{\rm Renyi} = \lim_{n\to1}
-\frac{i}{2(n-1)} \frac{{\cal A}(\Sg)}{H^{d-2}} F(n) \, ,
\end{align}
where 
\beq
\label{eq:Fw}
F(n) = \int_{\de^2}^\infty \frac{{\rm d} s}{s}  
\int_0^\infty {\rm d}\rho \rho (1+ H^2 \rho^2)^{\frac{d-3}{2}}
\int_{\G} {\rm d} \om \cot\left(\frac{\om}{2n}\right) K_1(\mu;s) \, ,
\eeq
$\cosh^2\left(\frac{H\mu}{2}\right) = 1 + (H\rho)^2 \sin^2(\om/2)$ 
and $K_1(\mu;s)$ is the EAdS heat-kernel on manifold ${\cal M}$ \cite{Camporesi:1990wm}. 
This is the exact expression for the EE and Renyi entropy, and it goes 
as the area of the surface $\Sg$ which is denoted above by ${\cal A}(\Sg)$. 
 
The integral stated in eq. (\ref{eq:Fw}) however is not easy to compute. 
It consists of three integrals: one over $\rho$ which can be traded to integral 
over $\mu$ using the relation $\cosh^2\left(\frac{H\mu}{2}\right) 
= 1 + (H\rho)^2 \sin^2(\om/2)$, the second is a 
contour integral over $\om$ for contour $\G$ while the last integral is 
over $s$. In the former one can do transformation of variables from 
$\rho$ to $\mu$. This leads to following expression for $F(n)$.
\beq
\label{eq:Fn_mu}
F(n) = \frac{1}{4H}  \int_{\de^2}^\infty \frac{{\rm d} s}{s}
\int_{\G} {\rm d} \om \frac{\cot\left(\om/2n\right)}{\sin^2(\om/2)}
\int_{0}^\infty {\rm d}\mu \sinh(H\mu) 
\left[1 + \frac{\sinh^2\left(H\mu/2\right)}{\sin^2\left(\om/2\right)} \right]^{(d-3)/2} K_1^d(\mu;s) \, ,
\eeq
where $K_1^d(\mu;s)$ is the heat-kernel on the AdS metric which is 
well known from the literature \cite{Camporesi:1990wm}. This integral is not 
easy to perform in arbitrary dimensions, however in odd dimensions 
the expression in square brackets becomes a polynomial in 
$\sinh^2(H\mu/2)$ when expanded. This offers some simplicity in 
obtaining expression for entropy in odd dimensions. 

%%%%%%%%%%%%%%%%%%%%%%%%%%%%%%%%%%%%%%%%%%%%%%%
\subsection{Non-local heat-kernel and effective action}
\label{NLHK}
%%%%%%%%%%%%%%%%%%%%%%%%%%%%%%%%%%%%%%%%%%%%%%%

The quantity that enters the computation of the entropy is the effective 
action. In this sun-section we will compute the effective action of the theory 
considered here by making use of the heat-kernel on the AdS. For the free 
non-local theory considered in eq. (\ref{eq:NLact}) (with the negative $-\lam^2$)
the effective action is given by,
\begin{align}
\label{eq:EAWNL}
W &= - \log Z
= \frac{1}{2} \log \det \left(
-\Box + m^2 + \frac{\lam^2}{-\Box} \right)
= \frac{1}{2} \log \det \left[
(-\Box)^{-1} \left( \Box^2 - m^2 \Box + \lam^2 
\right) \right] \, ,
\notag \\
&= -\frac{1}{2} \log \det(-\Box) 
+ \frac{1}{2} \log \det(-\Box + r_{-}^2)
+ \frac{1}{2} \log \det(-\Box + r_{+}^2) \, ,
\end{align}
where in obtaining last line we have used eq. (\ref{eq:opProd})
and, $r_{-}^2$ and $r_{+}^2$ are given in eq. (\ref{eq:RpRm}) 
respectively. Each of the piece 
in the last line of eq. (\ref{eq:EAWNL}) can be expressed in terms of 
heat-kernel of $\Box$-operator on AdS which gives the following expression 
for the effective action of our theory 
\beq
\label{eq:W1NLEA}
W_1 = \lim_{x\to x^\prime} \frac{1}{2} \int_{\de^2}^\infty \left(-\frac{{\rm d}s}{s} \right)
\left(e^{-s r_{+}^2} + e^{-s r_{-}^2} - 1 \right) {\rm Tr} K_0(x,x^\prime;s) \, ,
\eeq
where $K_0(x,x^\prime;s) = \bra{x} e^{-s(-\Box)}\ket{x^\prime}$ is the heat-kernel 
of the $\Box$. The eq. (\ref{eq:W1NLEA}) tell us that $K_1$ appearing in 
eq. (\ref{eq:Fn_mu}) is given by 
\beq
\label{eq:KNL}
K_1(x,x^\prime;s) = \left(e^{-s r_{+}^2} + e^{-s r_{-}^2} - 1 \right) K_0(x,x^\prime;s) \, .
\eeq
In the case when non-locality strength goes to zero ($\lam\to0$) 
one recovers the heat-kernel of the local massive scalar field operator
as $r_{-}^2 \to 0$. The heat-kernel $K_0$ on AdS background is 
known in closed form exactly \cite{Camporesi:1990wm} and is different 
in odd and even dimensions. For the AdS geometry it is given by,
\begin{align}
\label{eq:HKadsodd}
K_0^{\rm odd} &= \left(\frac{-1}{2\pi}\right)^{\frac{d-1}{2}} (4\pi s)^{-1/2} 
\left[ \frac{H}{\sinh(H\mu)} \frac{\pt}{\pt\mu} \right]^{\frac{d-1}{2}}
\exp\left[-\frac{\mu^2}{4s} - \frac{(d-1)^2H^2s}{4}\right] \, ,
\\
\label{eq:HKadseven}
K_0^{\rm even} &=
\left(\frac{-1}{2\pi}\right)^{\frac{d-2}{2}} \frac{\sqrt{2} 
e^{-(d-1)^2H^2s/4}}{(4\pi s)^{3/2}}
\left[ \frac{H}{\sinh(H\mu)} \frac{\pt}{\pt\mu} \right]^{\frac{d-2}{2}}
\int_{H\mu}^\infty {\rm d}\mu^\prime 
\frac{H^2\mu^\prime e^{-\mu^{\prime2}/4s}}
{\sqrt{\cosh(H\mu^\prime) - \cosh(H\mu)}} \, .
\end{align}
These can then be plugged in eq. (\ref{eq:KNL}) to obtain 
the expression for $K_1$.

%%%%%%%%%%%%%%%%%%%%%%%%%%%%%%%%%%%%%%%%%%%%%%%
\subsection{$AdS_3$}
\label{ads3}
%%%%%%%%%%%%%%%%%%%%%%%%%%%%%%%%%%%%%%%%%%%%%%%

We first consider the case of three dimensional AdS space. By making use of the 
heat-kernel for $K_0$ given in eq. (\ref{eq:HKadsodd}) and using the 
expression for $K_1$ from eq. (\ref{eq:KNL}) one can obtain $K_1$ 
for three-dimensional AdS to be,
\beq
\label{eq:ads3K1}
K_1^3(\mu;s) = \frac{H\mu\left[2e^{-m^2s/2} \cosh(s\sqrt{m^4 - 4\lam^2}/2)-1\right]
\exp\left(- H^2s -\mu^2/4s\right)}
{(4\pi s)^{3/2} \sinh(H\mu)} \, .
\eeq
This is the kernel for the non-local theory considered here. 
The kernel for the local massive theory can be obtained 
by taking the limit $\lam\to0$ which matches with the 
expression in \cite{Camporesi:1990wm,Giombi:2008vd,David:2009xg}. 
Using the Sommerfeld formula stated in eq. (\ref{eq:HKmodi}) one can 
compute the kernel on ${\cal M}^n$. In the computation of entropy 
we are however interested in quantity $F(n)$ which is mentioned 
in eq. (\ref{eq:Fn_mu}) and uses the heat-kernel. In $3d$ 
both the entanglement and Renyi (in the limit $n\to1$) entropy are same. 
It is given by,
\beq
\label{eq:EERen3d}
S_{\rm EE} = S_{\rm Renyi}
= \frac{2{\cal A}^1(\Sg)}{3\sqrt{\pi}}\left[
\frac{1}{\de} + \sqrt{\pi} \left(H -\sqrt{H^2 + r_{-}^2} - \sqrt{H^2 + r_{+}^2}\right)
+ \cdots 
\right] \, ,
\eeq
where ${\cal A}^1(\Sg)$ is the area of the surface in three-dimensions. 
The leading term is the UV divergent piece which diverges in the limit 
$\de\to0$ while the second term is the finite term. Terms of order 
${\cal O}(\de)$ have been ignored as they don't survive in the UV limit. 
The leading divergent piece is independent of the mass or non-locality 
present in the theory, however the finite piece contains information about 
both mass and non-locality. This finite piece is analytic in $m$ and $\lam$, 
and is always real. In particular one can write the finite piece in $3D$ as follows,
\begin{align}
\label{eq:finiteEE3D}
S_{EE}^{\rm finite} &= \frac{2{\cal A}^1(\Sg)}{3}
\left(H -\sqrt{H^2 + r_{-}^2} - \sqrt{H^2 + r_{+}^2}\right)
\notag \\
&= \frac{2{\cal A}^1(\Sg)}{3}
\left(H - \sqrt{u} e^{-v/2} - \sqrt{u} e^{v/2} \right) 
= \frac{2{\cal A}^1(\Sg)}{3}\left(H - 2 \sqrt{u} \cosh(v/2) \right) \, ,
\end{align}
where $u = \sqrt{H^4 + m^2 H^2 + \lam^2}$ 
and $\tanh v = \sqrt{m^4 - 4\lam^2}/(2H^2+m^2)$. If $v$ changes sign 
or becomes imaginary, still the entropy will remain real except 
in later case $\cosh(v/2) \to \cos(v/2)$. The entropy then 
has a oscillating part. In the locality limit 
($\lam\to0$): $r_{-}^2\to0$, $r_{+}^2\to m^2$ one obtains 
the case of massive local scalar field theory. 
The other interesting limit is the massless limit $m\to0$: 
$r_{-}^2\to-i \lam$, $r_{+}^2\to i\lam$, $v$ becomes imaginary
and $u \to \sqrt{H^4 + \lam^2}$. In the massless limit the entropy 
acquires a oscillating part. This oscillation is entirety due to 
the presence of non-locality. 
Also, as the end result is analytic in $m$ and $\lam$ therefore the change 
of parameter $\lam\to i \lam$ remains well-defined.  

%%%%%%%%%%%%%%%%%%%%%%%%%%%%%%%%%%%%%%%%%%%%%%%
\subsection{Generic Odd dimension}
\label{ads5}
%%%%%%%%%%%%%%%%%%%%%%%%%%%%%%%%%%%%%%%%%%%%%%%

In generic odd dimensions one can compute entanglement and Renyi 
entropy using the heat-kernel given in eq. (\ref{eq:HKadsodd}). This can then 
be plugged in eq. (\ref{eq:KNL}) to obtain the heat-kernel for the non-local 
theory considered here and consequently the effective action of the non-local theory
following eq. (\ref{eq:W1NLEA}). These are then utilised in eq. (\ref{eq:Fn_mu})
which immediately gives the entropy. The whole complications resides in 
computation of $F(n)$ in generic odd dimensions. 

On applying a derivative with respect to $\mu$ on $K_0$ it is seen 
that it follows a recursive relation, which is true for any dimensions. 
This recursion in $K_0$ also implies a recursion in $K_1$ given in 
eq. (\ref{eq:KNL}). 
\beq
\label{eq:HKrecur}
K_1^{d+2} (\mu,s) = 
- \frac{e^{-d H^2 s} H}{2 \pi \sinh(H\mu)} \frac{\pt}{\pt \mu} K_1^d(\mu,s) \, .
\eeq
This recursive relation is useful in obtaining some the terms of the entropy 
in generic dimensions. In general dimensions we can define the following 
quantity 
\beq
\label{eq:Ids}
I(d,s) = \int_{0}^\infty {\rm d}\mu \sinh(H\mu) 
\left[1 + \frac{\sinh^2\left(H\mu/2\right)}{\sin^2\left(\om/2\right)} \right]^{(d-3)/2} K_1^d(\mu;s) \, .
\eeq
Using the recursive property of $K_1^d(\mu,s)$ stated in eq. (\ref{eq:HKrecur}), 
it is noticed that $I(d,s)$ satisfies a corresponding recursive relation which is given by,
\beq
\label{eq:Idrecur}
I(d,s) = \frac{H}{2\pi} e^{-(d-2)H^2s} \left[
K_1^{d-2}(0;s) + \frac{(d-3)H}{4 \sin^2(\om/2)} I(d-2,s)
\right] \, .
\eeq
By making use of such relations it is possible to compute $I(d,s)$ recursively. 
In particular in space-time with odd $d$ it contains only a finite 
number of terms in $I(d,s)$. Such recursive relation offers an efficient 
algorithm in computing $I(d,s)$. In appendix \ref{Idsodd} we write $I(d,s)$ for 
some of the odd dimensions. 

Once we have computed $I(d,s)$ then the next integration is performed 
over $\om$ which is a contour integral. This involves computing residues. 
In the appendix \ref{Idsodd} we write the expression for result of 
the contour integration for some of the first few dimensions. Finally one needs 
to perform the integration over $s$ whose limit ranges from $\de^2$
and $\infty$. This integration reveals a Laurent series structure for 
$F(n)$ as the $s$-integration will have UV-divergences. 
\beq
\label{eq:FnLau}
F(n) = \sum_{k=-(d-2)}^{\infty} F_k(n) \de^k \, ,
\eeq
where $F_k(n)$ depend of the various parameters of the theory. 
The coefficient of negative powers of $\de$ are the UV divergent 
pieces. The leading divergence comes entirely from the local 
physics and doesn't depend either on mass or non-locality strength $\lam$. 
Sub-leading divergences however depend on mass and non-locality. 
In the appendix \ref{Idsodd} we have written $F_k(n)$ 
corresponding to UV divergence for some odd dimensions. 
In general their expression can be computed iteratively. The coefficient 
$F_0(n)$ is universal and UV finite. This contributes a universal 
piece to the entropy. For some of the odd dimensions 
it is given by,
\begin{align}
\label{eq:Fd30n}
F^3_0(n) &= -\frac{i H \left(n^2-1\right)\sqrt{2} }{3 n} \biggl[\sqrt{2 H^2-\sqrt{m^4-4 \lambda ^2}+m^2}
\notag \\
& +\sqrt{2 H^2+\sqrt{m^4-4 \lambda ^2}+m^2} - \sqrt{2} H \biggr] \, ,
\\
\label{eq:Fd50n}
F^5_0(n) &= -\frac{i H \left(n^2-1\right)}{180 \pi  n^3} 
\biggl[(116 H^3 n^2-4 H^3+\sqrt{2} H^2 \sqrt{8 H^2-\sqrt{m^4-4 \lambda ^2}+m^2}
\notag \\
&
+\sqrt{2} H^2
\sqrt{8 H^2+\sqrt{m^4-4 \lambda ^2}+m^2}-29 \sqrt{2} H^2 n^2 \sqrt{8 H^2-\sqrt{m^4-4 \lambda ^2}+m^2}
\notag \\
&
-29 \sqrt{2} H^2 n^2 \sqrt{8 H^2+\sqrt{m^4-4 \lambda ^2}+m^2}-5 \sqrt{2} m^2 n^2 
\sqrt{8 H^2-\sqrt{m^4-4 \lambda ^2}+m^2}
\notag \\
&+5 \sqrt{2} n^2 \sqrt{m^4-4 \lambda ^2} \sqrt{8 H^2-\sqrt{m^4-4 \lambda ^2}+m^2}
-5 \sqrt{2} m^2 n^2 \sqrt{8 H^2+\sqrt{m^4-4 \lambda^2}+m^2}
\notag \\
& -5 \sqrt{2} n^2 \sqrt{m^4-4 \lambda ^2} \sqrt{8 H^2+\sqrt{m^4-4 \lambda ^2}+m^2}\biggr] \, .
\end{align}
The interesting observation that should be made is 
that if we make a series of transformations: 
\beq
\label{eq:scale}
m\to \g m_*\ , \hspace{3mm} H \to \g H_* \, ,
\hspace{3mm} 
\lam \to \g^2 \lam_* \, .
\eeq
(where $m_*$, $H_*$ and $\lam_*$ are dimensionless 
parameters) then in odd dimensions the divergent contributions to the entropy is always 
odd in $\g$ while the universal finite term is always even in $\g$. This simple 
observation is very powerful in isolating the universal part of the entropy. 
In odd dimensional $d$ the universal finite piece is always coefficient of 
$\g^{d-1}$. In the following we will be interested in determining this 
universal piece. We will obtain this universal piece using green's function. 

%%%%%%%%%%%%%%%%%%%%%%%%%%%%%%%%%%%%%%%%%%%%%%%
\subsection{Universal finite piece from green's function}
\label{eq:greenUni}
%%%%%%%%%%%%%%%%%%%%%%%%%%%%%%%%%%%%%%%%%%%%%%%

In this subsection we will derive an expression for the universal piece 
of the entropy by making use of green's function. It should be observed that 
in odd dimensions the universal piece is the UV finite part which 
arises for $\de=0$. Also, after scaling (\ref{eq:scale}) in odd dimensions 
the divergent part is odd while the universal part even. As a result 
the quantity $(S_{\g} + S_{-\g})/2$ being even in $\g$ will directly give the universal 
piece due to the cancellation of the all the UV divergent terms. In the following 
we will be interested in computing this quantity. This quantity can actually 
be written as,
\beq
\label{eq:SgamInt}
\frac{1}{2} \int_{0}^{\g} {\rm d} \g^\prime \frac{\pt S}{\pt \g^\prime}
- \frac{1}{2} \int_{-\g}^{0} {\rm d} \g^\prime \frac{\pt S}{\pt \g^\prime}
= \frac{S(\g) + S(-\g)}{2} - S(0) \, .
\eeq 
The quantity $S(0)$ is the entropy computed for $\g=0$, which results 
in $S(0)=0$. This immediately brings the focus on the computation of 
$\pt S/\pt \g$, which can be obtained from green's function of the theory. 
The effective action of the theory in terms of heat-kernel is 
written in eq. (\ref{eq:W1NLEA}). In this if we do the scaling of parameters 
as indicated in eq. (\ref{eq:scale}) then the effective action is given by,
\beq
\label{eq:scaleW1}
W_1 = \lim_{x\to x^\prime} \int {\rm d}x \sqrt{g} \int_0^\infty \left(-\frac{{\rm d}s}{s}\right)
\left(e^{-s\g^2 r_{-}^{*2}} + e^{-s \g^2 r_{+}^{*2}} -1 \right) {\rm Tr} K_0(x,x^\prime;s) \, ,
\eeq
where $K_0(x,x^\prime;s)$ is the heat-kernel for the $\Box$ operator as before.
On taking derivative with respect to $\g$ we get,
\beq
\label{eq:Wdergam}
\frac{\pt W_1}{\pt \g} = 2 \g \lim_{x\to x^\prime} \int {\rm d}x \sqrt{g} \int_0^\infty 
{\rm d}s \left(e^{-s\g^2 r_{-}^{*2}} + e^{-s \g^2 r_{+}^{*2}} \right) {\rm Tr} K_0(x,x^\prime;s) \, .
\eeq
From this one can quickly recognise the form of the green's function 
appearing in terms of heat-kernel, which is the standard definition 
of green's function in terms of heat-kernel \cite{DeWitt:1965jb}. 
\beq
\label{eq:G1G2HK}
G_{+}(x,x^\prime) = \int_{0}^\infty {\rm d}s e^{-sr_{+}^{2}} K_0(x,x^\prime;s)\, ,
\hspace{5mm}
G_{-}(x,x^\prime) = \int_{0}^\infty {\rm d}s e^{-sr_{-}^{2}} K_0(x,x^\prime;s) \, .
\eeq
The green's function $G_{-}$ and $G_{+}$ have also been previously obtained 
by solving hyper-geometric differential equation. They satisfy the 
equations eq. (\ref{eq:G1G2eq}) and have solution given in 
eq. (\ref{eq:GNLmdSform}) with $q_{-}$ and $q_{+}$ given in 
eq. (\ref{eq:Q1Q2}). These are known exactly in AdS background
as has been shown in section \ref{ads} in closed form and is also 
well known from literature \cite{Allen:1985wd,Burgess:1984ti,Caldarelli:1998wk,Gubser:2002zh}. 
In the following we will be using these to compute the expression for $\pt S/\pt \g$. 

The green's function $G_{-}$ and $G_{+}$ can be suitably generalised 
to obtain on the $n$-copies of AdS following Sommerfeld formula 
which leads to following expression 
for the green's function on the $n$-manifold ${\cal M}^n$.
\bea
\label{eq:greenMod}
(G^\pm_n - G^\pm_1)(\phi,\phi^\prime) 
= \frac{i}{4\pi n} \int_{\G} {\rm d} \om \cot\left(\frac{\om}{2n}\right)
G^\pm_1(\phi-\phi^\prime+\om)
\eea
This will give corresponding effective action on ${\cal M}^n$ following eq. (\ref{eq:Wdergam}). 
\bea
\label{eq:Wnder}
&&
\frac{\pt W_n}{\pt \g} = 2\g {\rm Tr} \, \left[
G^{+}_n(x,x;m^2) + G^{-}_n(x,x;m^2)
\right]
= 2 \g H^{3-d} \int_0^{2\pi n/H} {\rm d}\tau_E 
\notag \\
&&
\times
\int_{H^{d-2}} {\rm d}V_{d-2} 
\int_0^\infty {\rm d}\rho \, \rho(1 + H^2 \rho^2)^{\frac{d-3}{2}} \, \left[
G^{+}_n(x,x;m^2) + G^{-}_n(x,x;m^2)
\right] \, .
\eea
where $G^\pm_n(x,x^\prime)$ is obtained in section \ref{ads}.
This allows us to translate the heat-kernel and the green's function of scalar in EAdS 
in to heat-kernel and green's function on metric eq. (\ref{eq:adstopobh}) respectively. 
These will be used in the computation of entanglement entropy (and Renyi entropy). 
In this case we have
\begin{align}
\label{eq:EEderexp}
&
\frac{\pt S_{\rm EE}}{\pt \g} =  \lim_{n\to1}
-\frac{i}{2} \frac{{\cal A}(\Sg)}{H^{d-2}}  
(n\pt_n-1) F^G(n) \, , 
\\
\label{eq:Renyiderform}
& 
\frac{\pt S_{\rm Renyi}}{\pt \g}  =  \lim_{n\to1}
-\frac{i}{2(n-1)} \frac{{\cal A}(\Sg)}{H^{d-2}} F^G(n) \, ,
\end{align}
where 
\beq
\label{eq:FGw}
F^G(n) = 2\g \int_0^\infty {\rm d}\rho \rho (1+ H^2 \rho^2)^{\frac{d-3}{2}}
\int_{\G} {\rm d} \om \cot\left(\frac{\om}{2n}\right) \left[
G^{+}_1(\mu) + G^{-}_1(\mu)
\right] \, ,
\eeq
$\cosh^2\left(\frac{H\mu}{2}\right) =z= 1 + (H\rho)^2 \sin^2(\om/2)$ 
(as before) and $G^\pm_1(\mu)$ is the EAdS green's function 
computed previously using differential equation. 
A change of variable which leads ${\rm d}z = 2 H^2 \rho \sin^2(\om/2) {\rm d}\rho$
with the parameter $z$ range $1\leq z\leq\infty$. Then we choose to first 
perform the integration over $\om$ and then over $z$. This interchange of 
order of integration leads to,
\begin{align}
\label{eq:Fw_Z}
F^G(n) = 
\frac{\g}{H^2} \int_1^\infty {\rm d} z 
\left[G^{+}_1(z) + G^{-}_1(z) \right] \int_{\G} {\rm d} \om 
\frac{\cot\left(\om/2n\right)}{\sin^2\left(\om/2\right)}
\left[1 + \frac{z-1}{\sin^2\left(\om/2\right)} \right]^{(d-3)/2} \, . 
\end{align}
In generic odd dimensions $d=2 j+1$ (where $j \in \mathbb{I}$) one has
the following polynomial expansion with finite number of terms in expansion. 
\beq
\label{eq:binom}
\left[1 + \frac{z-1}{\sin^2\left(\om/2\right)} \right]^{j-1} 
= \sum_{k=0}^{j-1} \frac{(j-1)!}{k! (j-k-1)!} 
\left(\frac{z-1}{\sin^2\left(\om/2\right)}
\right)^k \, .
\eeq
In even dimensions this binomial expansion has infinite number of 
terms in series. 
In odd dimensional AdS, using this expansion one can perform 
the $\om$-integration, which leads to following expression for 
$F^G(n) = (\g/H^2) \sum_k F^G_k$ with $F^G_k$'s being function 
of $n$ given by
\beq
\label{eq:Fser}
F^G_k =  
\frac{\G(d-2) C_k(n)}{\G(k) \G(d-2+k)} 
\int_1^\infty {\rm d} z (z-1)^k \left[G^{+}_1(z) + G^{-}_1(z) \right] \, ,
\eeq
where the functions $C_k(n)$ are obtained during the 
process of computing contour integration. 
\beq
\label{eq:Ckn}
C_k(n) =  \int_{\G} {\rm d} \om \cot\left(\frac{\om}{2n}\right)
\left( \sin\left(\frac{\om}{2}\right) \right)^{-(2k+2)} \, .
\eeq
The functions $C_k(n)$ are coefficients containing information of 
the background involving conical singularity \cite{Dowker:1987mn,Fursaev:1994in}
and are rational functions in $n$. In the appendix \ref{Ckn} 
we write some of the first few $C_k$'s. 
The integration over $z$ as given in 
eq. (\ref{eq:Fser}) can be performed on {\it Mathematica}
giving a closed form expression in terms of 
hyper-geometric functions. This is given by
\begin{align}
\label{eq:Gint}
&
\int_1^\infty (z-1)^k \left[G^{+}_1(z) + G^{-}_1(z) \right]
= \G(k+1) \biggl[
\frac{q_{-}\G(a_{-}-k-1)}{\G(a_{-})} 
\notag \\
& \times 
{}_2F_1(a_{-}-k-1, a_{-}-c_{-}+1, a_{-}-b_{-}+1,1)
+ \frac{q_{+}\G(a_{+}-k-1)}{\G(a_{+})} 
\notag \\
&
\times {}_2F_1(a_{+}-k-1, a_{+}-c_{+}+1, a_{+}-b_{+}+1,1)
\biggr] \, ,
\end{align}
where we have used the structure of the green's function $G_{-}$ and $G_{+}$
on EAdS obtained previously, with $q_{-}$ and $q_{+}$ 
given in eq. (\ref{eq:Q1Q2}) respectively. 
The hyper-geometric function has a singularity at argument $z=1$, 
but for now we don't worry about it. At this point one can combine 
eq. (\ref{eq:Fser}), (\ref{eq:Ckn}) and (\ref{eq:Gint}) to obtain an 
expression for $F^G(n)$ given in eq. (\ref{eq:Fw_Z}). This is the main ingredient 
that is required for the computation of the entanglement or Renyi entropy. 
The function $F^G(n)$ is given by,
\begin{align}
\label{eq:Ffullform}
&
F^G(n) = \frac{\g}{H^2} \sum_{k=0}^{j-1}
\frac{\G(d-2) C_k(n)}{\G(k) \G(d-2+k)} \G(k+1) 
\times
\notag \\
&
\biggl[
\frac{q_{-} \G(a_{-}-k-1)}{\G(a_{-})} 
{}_2F_1(a_{-}-k-1, a_{-}-c_{-}+1, a_{-}-b_{-}+1,1)
\notag \\
& + \frac{q_{+} \G(a_{+}-k-1)}{\G(a_{+})} 
{}_2F_1(a_{+}-k-1, a_{+}-c_{+}+1, a_{+}-b_{+}+1,1)
\biggr] \, .
\end{align}
At this point one can do the scaling of parameters using eq. (\ref{eq:scale}).
From the resulting expression we compute the universal 
finite part of entropy following 
eq. (\ref{eq:EEderexp}) and (\ref{eq:Renyiderform}) by
integrating with respect to $\g$ and using eq. (\ref{eq:SgamInt}). 
At this point 
we notice that the $n$-derivate and the $n$-limit acts directly on the 
functions $C_k(n)$. This series has a finite number of terms and 
offers a closed form expression. In the computation of entropy this 
is the universal piece.

%%%%%%%%%%%%%%%%%%%%%%%%%%%%%%%%%%%%%%%%%%%%%%%
\section{Conclusions}
\label{conc}
%%%%%%%%%%%%%%%%%%%%%%%%%%%%%%%%%%%%%%%%%%%%%%%

In this  paper we have considered a non-local scalar field theory and investigated 
entanglement entropy in this theory in odd AdS spacetime. 

First, the Green function of the theory in flat and AdS space-time was
computed and analyzed. Using the trick of partial decomposition we wrote 
the propagator of non-local theory as a linear sum propagator of 
two local theories. The Green function so computed is well-behaved in 
ultraviolet (short-distances) and infrared (long-distances). The massless 
limit of the non-local propagator in AdS is well-defined. 

We then calculated the entanglement entropy in this theory with the purpose to investigate 
whether non-locality can play a significant role in affecting long-distance 
entanglement. We  used replica trick. For this we first computed the 
expression for the effective action of the non-local theory. This was then 
written in the language of heat-kernel on the AdS. 
Using Sommerfeld formula we then computed the heat-kernel 
and subsequently the effective action on the $n$-copies 
of the manifold ${\cal M}^n$. This allowed us to compute the 
entanglement and Renyi entropy via replica trick.
We computed the 
entropy in odd space-time dimensions. We found that the area-law 
is followed even for non-local theory. The leading term of the 
entropy is independent of any of the parameters of the theory. 
The sub-leading terms depends on mass an non-locality 
strength. This is expected in the sense that the leading term arises 
because of the ultralocal physics and hence doesn't contain any dependence 
on mass or non-locality. 

In odd-dimensions the entropy is seen to have a 
universal part which is UV finite. We computed the form of this universal part 
of the entropy in all odd dimensions using Green functions, which offer a quick and elegant way to compute 
universal contributions to entropy. 
This UV finite piece unlike the leading term has more structure 
and incorporate the effect from non-locality. 
In $AdS_{3}$, this universal part is seen to be oscillating 
in massless theories. This oscillation is entirely due to the 
presence of non-locality in the theory, which doesn't happen in case 
of local theories \cite{Sugishita:2016iel}, and can be seen 
as a novelty due to the presence of non-locality. 
The oscillation can be understood physically by expressing the non-local theory as a local one, 
where one can notice the presence of a tachyon or a complex-pole. 
This is only indicating that the non-local theory is an effective one in infrared 
regimes and cannot become a fundamental local theory.

%%%%%%%%%%%%%%%%%%%%%%%%%%%%%%%%%%%%%%%%%%%%%%%
\bigskip
\centerline{\bf Acknowledgements} 
%%%%%%%%%%%%%%%%%%%%%%%%%%%%%%%%%%%%%%%%%%%%%%%

NK is supported by the SERB National Postdoctoral Fellowship. 
GN is supported by ``Zhuoyue" Fellowship (ZYBH2018-03).

%%%%%%%%%%%%%%%%%%%%%%%%%%%%%%%%%%%%%%%%%%%%%%%
\begin{appendices}
%%%%%%%%%%%%%%%%%%%%%%%%%%%%%%%%%%%%%%%%%%%%%%%

%%%%%%%%%%%%%%%%%%%%%%%%%%%%%%%%%%%%%%%%%%%%%%%
\section{$I(d,s)$ in odd $d$}
\label{Idsodd}
%%%%%%%%%%%%%%%%%%%%%%%%%%%%%%%%%%%%%%%%%%%%%%%

Here we write the expression for $I(d,s)$ for some of the 
odd dimensions. $I(d,s)$ is given in eq. (\ref{eq:Ids}) and follows 
a recursive relation mentioned in eq. (\ref{eq:Idrecur}). 
\begin{align}
\label{eq:ids3}
I(3,s) &= \frac{H e^{-H^2 s} \left[2 e^{-s m^2/2} \cosh \left(\frac{1}{2} s \sqrt{m^4-4 \lambda ^2}\right)-1\right]}
{4 \pi ^{3/2} \sqrt{s}} \, ,
\\
\label{eq:ids5}
I(5,s) &= -\frac{H e^{-4 H^2 s} \left(H^2 s \csc ^2\left(\frac{\omega }{2}\right)+1\right) 
\left[1-2 e^{-m^2s/2} \cosh \left(\frac{1}{2} s \sqrt{m^4-4 \lambda^2}\right)\right]}
{16 \pi ^{5/2} s^{3/2}} \, ,
\\
\label{eq:ids7}
I(7,s) &= -\frac{He^{-9 H^2s}}{192 \pi ^{7/2} s^{5/2}} \left[1-2 e^{-m^2s/2} 
\cosh \left(\frac{1}{2} s \sqrt{m^4-4 \lambda ^2}\right)\right] 
\notag \\
& \times \left[
\frac{6H^4 s^2}{\sin^4(\om/2)} + \frac{6H^2s}{\sin^2(\om/2)} + 2H^2 s+3
\right] \, .
\end{align}
Once the $I(d,s)$ are known one can compute the contour integration over $\om$. 
The integral that we are interested in performing is
\beq
\label{eq:wintIds}
J(d,n,s) = \frac{i n}{(n^2-1)H} 
\int_{\G} {\rm d} \om \frac{\cot\left(\om/2n\right)}{\sin^2(\om/2)}
I(d,s) \, .
\eeq
For some of the first few odd dimensions they are given by, 
\begin{align}
\label{eq:Jd3}
J(3,n,s) &= \frac{e^{-H^2s} \left[1 - 2 e^{-m^2s/2}
\cosh\left(\frac{1}{2} s \sqrt{m^4-4 \lambda ^2}\right) \right]}{3 \sqrt{s\pi}} \, ,
\\
\label{eq:Jd5}
J(5,n,s) &=\frac{e^{-4H^2s} \left[1 - 2 e^{-m^2s/2}
\cosh\left(\frac{1}{2} s \sqrt{m^4-4 \lambda ^2}\right) \right]
\left[n^2 \left(11 H^2 s+15\right)+H^2 s\right]}
{180 n^2 (s\pi)^{3/2}} \, ,
\\
\label{eq:Jd7}
J(7,n,s) &= \frac{e^{-9H^2s} \left[1 - 2 e^{-m^2s/2}
\cosh\left(\frac{1}{2} s \sqrt{m^4-4 \lambda ^2}\right) \right]}
{15120 n^4 (s\pi)^{5/2}}
\notag \\
& \times \left[2 H^4 \left(191 n^4+23 n^2+2\right) s^2+42 H^2 n^2 \left(16
n^2+1\right) s+315 n^4\right]
\end{align}
From this one can compute $F(n)$ following eq. (\ref{eq:Fn_mu}) 
and the various divergent pieces of $F(n)$. Below we will mention 
divergent pieces of $F(n)$ for some odd dimensions. 
\begin{align}
\label{eq:Fd3m1}
& F^3_{-1} = \frac{2 i H \left(n^2-1\right)}{3 \sqrt{\pi } n} \, ,
\\
\label{eq:Fd5m13}
& F^5_{-3} = \frac{i H \left(n^2-1\right)}{18 \pi ^{3/2} n} \, ,
\hspace{5mm}
F^5_{-1} = -\frac{i H \left(n^2-1\right) \left(49 H^2 n^2-H^2+15 m^2 n^2\right)}{90 \pi ^{3/2} n^3} \, ,
\\
\label{eq:Fd7m5}
& F^7_{-5} = \frac{i H \left(n^2-1\right)}{120 \pi ^{5/2} n} \, ,
\hspace{5mm}
F^7_{-3} = -\frac{i H \left(n^2-1\right) \left(103 H^2 n^2-2 H^2+15 m^2 n^2\right)}{1080 \pi ^{5/2} n^3} \, ,
\notag \\
& F^7_{-1} = \frac{i H \left(n^2-1\right)}{15120 \pi ^{5/2} n^5} \bigl(14183 H^4 n^4
-664 H^4 n^2+8 H^4+4326 H^2 m^2 n^4-84 H^2 m^2 n^2
\notag \\
& +315 m^4 n^4-630 \lambda^2 n^4\bigr)
\end{align}
If we do scaling $m\to \g m_*$, $H \to \g H_*$
and $\lam \to \g^2 \lam_*$ (where $m_*$, $H_*$ and $\lam_*$ are dimensionless 
parameters) then it will be seen that all these are odd in $\g$.

%%%%%%%%%%%%%%%%%%%%%%%%%%%%%%%%%%%%%%%%%%%%%%%
\section{$C_k(n)$}
\label{Ckn}
%%%%%%%%%%%%%%%%%%%%%%%%%%%%%%%%%%%%%%%%%%%%%%%

Here we just mention some of the first few $C_k(n)$ following the 
residue theorem for the evaluation of contour integral.
\begin{align}
\label{eq:CK0}
& C_0(n) = \frac{4 i \pi  \left(n^2-1\right)}{3 n} \, ,
\\
\label{eq:CK1}
& C_1(n) = \frac{4 i \pi  (n^2-1) \left(11 n^2+1\right)}{45 n^3} \, ,
\\
\label{eq:CK2}
& C_2(n) = \frac{4 i \pi  (n^2-1)\left(191n^4 +23n^2 +2 \right)}{945 n^5} \, ,
\\
\label{eq:CK3}
&
C_3(n) = \frac{4 i \pi (n^2-1)(11n^2+1)(227n^4+10n^2+3)}{14175 n^7} \, ,
\\
\label{eq:CK4}
&
C_4(n) = \frac{4 i \pi (n^2-1)\left(14797 n^8+2125 n^6+321 n^4+35 n^2+2\right)}
{93555 n^9} \\
\label{eq:CK5}
&
C_5(n) = \frac{4 i \pi (n^2-1)}{638512875 n^{11}} 
\bigl(92427157 n^{10}+13803157 n^8+2295661 n^6
\notag \\
& +307961 n^4+28682 n^2+1382\bigr) \, .
\end{align}

%%%%%%%%%%%%%%%%%%%%%%%%%%%%%%%%%%%%%%%%%%%%%%%%%%%%%%%%%%%%%%%%%%%%

%%%%%%%%%%%%%%%%%%%%%%%%%%%%%%%%%%%%%%%%%%%%%%%
\end{appendices}
%%%%%%%%%%%%%%%%%%%%%%%%%%%%%%%%%%%%%%%%%%%%%%%

%%%%%%%%%%%%%%%%%%%%%%%%%%%%%%%%%%%%%%%%%%%%%%%%%%%%%%%%%%%%%%%%%%%%


\begin{thebibliography}{99} 
%
%
%\cite{Bombelli:1986rw}
\bibitem{Bombelli:1986rw} 
  L.~Bombelli, R.~K.~Koul, J.~Lee and R.~D.~Sorkin,
  %``A Quantum Source of Entropy for Black Holes,''
Phys.\ Rev.\ D {\bf 34}, 373 (1986).
doi:10.1103/PhysRevD.34.373
%%CITATION = doi:10.1103/PhysRevD.34.373;%%
  %886 citations counted in INSPIRE as of 03 Dec 2018


%\cite{Srednicki:1993im}
\bibitem{Srednicki:1993im} 
  M.~Srednicki,
  %``Entropy and area,''
Phys.\ Rev.\ Lett.\  {\bf 71}, 666 (1993)
doi:10.1103/PhysRevLett.71.666
[hep-th/9303048].
%%CITATION = doi:10.1103/PhysRevLett.71.666;%%
  %968 citations counted in INSPIRE as of 03 Dec 2018


%\cite{Casini:2009sr}
\bibitem{Casini:2009sr} 
  H.~Casini and M.~Huerta,
  %``Entanglement entropy in free quantum field theory,''
J.\ Phys.\ A {\bf 42}, 504007 (2009)
doi:10.1088/1751-8113/42/50/504007
[arXiv:0905.2562 [hep-th]].
%%CITATION = doi:10.1088/1751-8113/42/50/504007;%%
  %300 citations counted in INSPIRE as of 03 Dec 2018


%\cite{Ryu:2006ef}
\bibitem{Ryu:2006ef} 
  S.~Ryu and T.~Takayanagi,
  %``Aspects of Holographic Entanglement Entropy,''
JHEP {\bf 0608}, 045 (2006)
doi:10.1088/1126-6708/2006/08/045
[hep-th/0605073].
%%CITATION = doi:10.1088/1126-6708/2006/08/045;%%
  %1008 citations counted in INSPIRE as of 03 Dec 2018


%\cite{Faulkner:2013ana}
\bibitem{Faulkner:2013ana} 
  T.~Faulkner, A.~Lewkowycz and J.~Maldacena,
  %``Quantum corrections to holographic entanglement entropy,''
JHEP {\bf 1311}, 074 (2013)
doi:10.1007/JHEP11(2013)074
[arXiv:1307.2892 [hep-th]].
%%CITATION = doi:10.1007/JHEP11(2013)074;%%
  %241 citations counted in INSPIRE as of 03 Dec 2018


%\cite{Sugishita:2016iel}
\bibitem{Sugishita:2016iel} 
  S.~Sugishita,
  %``Entanglement entropy for free scalar fields in AdS,''
JHEP {\bf 1609}, 128 (2016)
doi:10.1007/JHEP09(2016)128
[arXiv:1608.00305 [hep-th]].
%%CITATION = doi:10.1007/JHEP09(2016)128;%%
  %2 citations counted in INSPIRE as of 03 Dec 2018


%\cite{Shiba:2013jja}
\bibitem{Shiba:2013jja} 
  N.~Shiba and T.~Takayanagi,
  %``Volume Law for the Entanglement Entropy in Non-local QFTs,''
JHEP {\bf 1402}, 033 (2014)
doi:10.1007/JHEP02(2014)033
[arXiv:1311.1643 [hep-th]].
%%CITATION = doi:10.1007/JHEP02(2014)033;%%
  %32 citations counted in INSPIRE as of 03 Dec 2018


%\cite{Karczmarek:2013xxa}
\bibitem{Karczmarek:2013xxa} 
  J.~L.~Karczmarek and C.~Rabideau,
  %``Holographic entanglement entropy in nonlocal theories,''
JHEP {\bf 1310}, 078 (2013)
doi:10.1007/JHEP10(2013)078
[arXiv:1307.3517 [hep-th]].
%%CITATION = doi:10.1007/JHEP10(2013)078;%%
  %29 citations counted in INSPIRE as of 03 Dec 2018

%\cite{Fischler:2013gsa}
\bibitem{Fischler:2013gsa} 
  W.~Fischler, A.~Kundu and S.~Kundu,
  %``Holographic Entanglement in a Noncommutative Gauge Theory,''
  JHEP {\bf 1401}, 137 (2014)
  doi:10.1007/JHEP01(2014)137
  [arXiv:1307.2932 [hep-th]].
  %%CITATION = doi:10.1007/JHEP01(2014)137;%%
  %35 citations counted in INSPIRE as of 10 Dec 2018


%\cite{Pang:2014tpa}
\bibitem{Pang:2014tpa} 
  D.~W.~Pang,
  %``Holographic entanglement entropy of nonlocal field theories,''
Phys.\ Rev.\ D {\bf 89}, no. 12, 126005 (2014)
doi:10.1103/PhysRevD.89.126005
[arXiv:1404.5419 [hep-th]].
%%CITATION = doi:10.1103/PhysRevD.89.126005;%%
  %6 citations counted in INSPIRE as of 03 Dec 2018


%\cite{Nesterov:2010yi}
\bibitem{Nesterov:2010yi} 
  D.~Nesterov and S.~N.~Solodukhin,
  %``Gravitational effective action and entanglement entropy in UV modified theories with and without Lorentz symmetry,''
  Nucl.\ Phys.\ B {\bf 842}, 141 (2011)
  doi:10.1016/j.nuclphysb.2010.08.006
  [arXiv:1007.1246 [hep-th]].
  %%CITATION = doi:10.1016/j.nuclphysb.2010.08.006;%%
  %34 citations counted in INSPIRE as of 10 Dec 2018


%\cite{Nesterov:2010jh}
\bibitem{Nesterov:2010jh} 
  D.~Nesterov and S.~N.~Solodukhin,
  %``Short-distance regularity of Green's function and UV divergences in entanglement entropy,''
  JHEP {\bf 1009}, 041 (2010)
  doi:10.1007/JHEP09(2010)041
  [arXiv:1008.0777 [hep-th]].
  %%CITATION = doi:10.1007/JHEP09(2010)041;%%
  %8 citations counted in INSPIRE as of 10 Dec 2018

%\cite{Solodukhin:2011gn}
\bibitem{Solodukhin:2011gn} 
  S.~N.~Solodukhin,
  %``Entanglement entropy of black holes,''
Living Rev.\ Rel.\  {\bf 14}, 8 (2011)
doi:10.12942/lrr-2011-8
[arXiv:1104.3712 [hep-th]].
%%CITATION = doi:10.12942/lrr-2011-8;%%
  %235 citations counted in INSPIRE as of 03 Dec 2018


%\cite{Modesto:2011kw}
\bibitem{Modesto:2011kw} 
  L.~Modesto,
  %``Super-renormalizable Quantum Gravity,''
Phys.\ Rev.\ D {\bf 86}, 044005 (2012)
doi:10.1103/PhysRevD.86.044005
[arXiv:1107.2403 [hep-th]].
%%CITATION = doi:10.1103/PhysRevD.86.044005;%%
  %267 citations counted in INSPIRE as of 03 Dec 2018


%\cite{Modesto:2017sdr}
\bibitem{Modesto:2017sdr} 
  L.~Modesto and L.~Rachwa?,
  %``Nonlocal quantum gravity: A review,''
Int.\ J.\ Mod.\ Phys.\ D {\bf 26}, no. 11, 1730020 (2017).
doi:10.1142/S0218271817300208
%%CITATION = doi:10.1142/S0218271817300208;%%
  %20 citations counted in INSPIRE as of 03 Dec 2018


%\cite{Maggiore:2016gpx}
\bibitem{Maggiore:2016gpx} 
  M.~Maggiore,
  %``Nonlocal Infrared Modifications of Gravity. A Review,''
Fundam.\ Theor.\ Phys.\  {\bf 187}, 221 (2017)
doi:10.1007/978-3-319-51700-116
[arXiv:1606.08784 [hep-th]].
%%CITATION = doi:10.1007/978-3-319-51700-1_16;%%
  %31 citations counted in INSPIRE as of 03 Dec 2018


%\cite{Narain:2017twx}
\bibitem{Narain:2017twx} 
  G.~Narain and T.~Li,
  %``Ultraviolet complete dark energy model,''
Phys.\ Rev.\ D {\bf 97}, no. 8, 083523 (2018)
doi:10.1103/PhysRevD.97.083523
[arXiv:1712.09054 [hep-th]].
%%CITATION = doi:10.1103/PhysRevD.97.083523;%%
  %3 citations counted in INSPIRE as of 03 Dec 2018


%\cite{Narain:2018hxw}
\bibitem{Narain:2018hxw} 
  G.~Narain and T.~Li,
  %``Non-Locality and Late-Time Cosmic Acceleration from an Ultraviolet Complete Theory,''
Universe {\bf 4}, no. 8, 82 (2018)
doi:10.3390/universe4080082
[arXiv:1807.10028 [hep-th]].
%%CITATION = doi:10.3390/universe4080082;%%


%\cite{Kajuri:2017jmy}
\bibitem{Kajuri:2017jmy} 
  N.~Kajuri,
  %``Unruh Effect in nonlocal field theories,''
Phys.\ Rev.\ D {\bf 95}, no. 10, 101701 (2017)
doi:10.1103/PhysRevD.95.101701
[arXiv:1704.03793 [gr-qc]].
%%CITATION = doi:10.1103/PhysRevD.95.101701;%%
  %6 citations counted in INSPIRE as of 03 Dec 2018


%\cite{Kajuri:2018myh}
\bibitem{Kajuri:2018myh} 
  N.~Kajuri and D.~Kothawala,
  %``Hawking radiation in Non local field theories,''
arXiv:1806.10345 [gr-qc].
%%CITATION = ARXIV:1806.10345;%%


%\cite{Kajuri:2018wow}
\bibitem{Kajuri:2018wow} 
  N.~Kajuri and G.~Narain,
  %``Boundary dual of a non local theory,''
  arXiv:1812.00946 [hep-th].
  %%CITATION = ARXIV:1812.00946;%%


%\cite{Avis:1977yn}
\bibitem{Avis:1977yn} 
  S.~J.~Avis, C.~J.~Isham and D.~Storey,
  %``Quantum Field Theory in anti-De Sitter Space-Time,''
Phys.\ Rev.\ D {\bf 18}, 3565 (1978).
doi:10.1103/PhysRevD.18.3565
%%CITATION = doi:10.1103/PhysRevD.18.3565;%%
  %365 citations counted in INSPIRE as of 03 Dec 2018


%\cite{Allen:1985wd}
\bibitem{Allen:1985wd} 
  B.~Allen and T.~Jacobson,
  %``Vector Two Point Functions in Maximally Symmetric Spaces,''
Commun.\ Math.\ Phys.\  {\bf 103}, 669 (1986).
doi:10.1007/BF01211169
%%CITATION = doi:10.1007/BF01211169;%%
  %245 citations counted in INSPIRE as of 03 Dec 2018


%\cite{Burgess:1984ti}
\bibitem{Burgess:1984ti} 
  C.~P.~Burgess and C.~A.~Lutken,
  %``Propagators and Effective Potentials in Anti-de Sitter Space,''
Phys.\ Lett.\  {\bf 153B}, 137 (1985).
doi:10.1016/0370-2693(85)91415-7
%%CITATION = doi:10.1016/0370-2693(85)91415-7;%%
  %140 citations counted in INSPIRE as of 03 Dec 2018


%\cite{Caldarelli:1998wk}
\bibitem{Caldarelli:1998wk} 
  M.~M.~Caldarelli,
  %``Quantum scalar fields on anti-de Sitter space-time,''
Nucl.\ Phys.\ B {\bf 549}, 499 (1999)
doi:10.1016/S0550-3213(99)00137-6
[hep-th/9809144].
%%CITATION = doi:10.1016/S0550-3213(99)00137-6;%%
  %29 citations counted in INSPIRE as of 03 Dec 2018


%\cite{Gubser:2002zh}
\bibitem{Gubser:2002zh} 
  S.~S.~Gubser and I.~Mitra,
  %``Double trace operators and one loop vacuum energy in AdS / CFT,''
Phys.\ Rev.\ D {\bf 67}, 064018 (2003)
doi:10.1103/PhysRevD.67.064018
[hep-th/0210093].
%%CITATION = doi:10.1103/PhysRevD.67.064018;%%
  %85 citations counted in INSPIRE as of 03 Dec 2018

%\cite{Narain:2018rif}
\bibitem{Narain:2018rif} 
  G.~Narain and N.~Kajuri,
  %``Non-local scalar field on deSitter and its infrared behaviour,''
  arXiv:1812.00947 [hep-th].
  %%CITATION = ARXIV:1812.00947;%%


%\cite{Callan:1994py}
\bibitem{Callan:1994py} 
  C.~G.~Callan, Jr. and F.~Wilczek,
  %``On geometric entropy,''
Phys.\ Lett.\ B {\bf 333}, 55 (1994)
doi:10.1016/0370-2693(94)91007-3
[hep-th/9401072].
%%CITATION = doi:10.1016/0370-2693(94)91007-3;%%
  %509 citations counted in INSPIRE as of 03 Dec 2018


%\cite{Holzhey:1994we}
\bibitem{Holzhey:1994we} 
  C.~Holzhey, F.~Larsen and F.~Wilczek,
  %``Geometric and renormalized entropy in conformal field theory,''
Nucl.\ Phys.\ B {\bf 424}, 443 (1994)
doi:10.1016/0550-3213(94)90402-2
[hep-th/9403108].
%%CITATION = doi:10.1016/0550-3213(94)90402-2;%%
  %717 citations counted in INSPIRE as of 03 Dec 2018


%\cite{Calabrese:2004eu}
\bibitem{Calabrese:2004eu} 
  P.~Calabrese and J.~L.~Cardy,
  %``Entanglement entropy and quantum field theory,''
J.\ Stat.\ Mech.\  {\bf 0406}, P06002 (2004)
doi:10.1088/1742-5468/2004/06/P06002
[hep-th/0405152].
%%CITATION = doi:10.1088/1742-5468/2004/06/P06002;%%
  %879 citations counted in INSPIRE as of 03 Dec 2018


%\cite{Calabrese:2009qy}
\bibitem{Calabrese:2009qy} 
  P.~Calabrese and J.~Cardy,
  %``Entanglement entropy and conformal field theory,''
J.\ Phys.\ A {\bf 42}, 504005 (2009)
doi:10.1088/1751-8113/42/50/504005
[arXiv:0905.4013 [cond-mat.stat-mech]].
%%CITATION = doi:10.1088/1751-8113/42/50/504005;%%
  %465 citations counted in INSPIRE as of 03 Dec 2018


%\cite{Dowker:1977zj}
\bibitem{Dowker:1977zj} 
  J.~S.~Dowker,
  %``Quantum Field Theory on a Cone,''
J.\ Phys.\ A {\bf 10}, 115 (1977).
doi:10.1088/0305-4470/10/1/023
%%CITATION = doi:10.1088/0305-4470/10/1/023;%%
  %139 citations counted in INSPIRE as of 03 Dec 2018


%\cite{Dowker:1987mn}
\bibitem{Dowker:1987mn} 
  J.~S.~Dowker,
  %``Casimir Effect Around a Cone,''
Phys.\ Rev.\ D {\bf 36}, 3095 (1987).
doi:10.1103/PhysRevD.36.3095
%%CITATION = doi:10.1103/PhysRevD.36.3095;%%
  %127 citations counted in INSPIRE as of 03 Dec 2018


%\cite{Fursaev:1994in}
\bibitem{Fursaev:1994in} 
  D.~V.~Fursaev,
  %``Spectral geometry and one loop divergences on manifolds with conical singularities,''
Phys.\ Lett.\ B {\bf 334}, 53 (1994)
doi:10.1016/0370-2693(94)90590-8
[hep-th/9405143].
%%CITATION = doi:10.1016/0370-2693(94)90590-8;%%
  %80 citations counted in INSPIRE as of 03 Dec 2018
%
%
%\cite{Sommerfeld:1897}  
\bibitem{Sommerfeld:1897} 
A.~Sommerfeld, 
``\"{U}ber verzweigte Potentiale im Raum,''
Proc. London Math. Soc. (1896)   s1-28  (1):  395-429. 
doi: 10.1112/plms/s1-28.1.395 
%
%
%\cite{Casini:2011kv}
\bibitem{Casini:2011kv} 
  H.~Casini, M.~Huerta and R.~C.~Myers,
  %``Towards a derivation of holographic entanglement entropy,''
JHEP {\bf 1105}, 036 (2011)
doi:10.1007/JHEP05(2011)036
[arXiv:1102.0440 [hep-th]].
%%CITATION = doi:10.1007/JHEP05(2011)036;%%
  %617 citations counted in INSPIRE as of 03 Dec 2018


%\cite{Hung:2014npa}
\bibitem{Hung:2014npa} 
  L.~Y.~Hung, R.~C.~Myers and M.~Smolkin,
  %``Twist operators in higher dimensions,''
JHEP {\bf 1410}, 178 (2014)
doi:10.1007/JHEP10(2014)178
[arXiv:1407.6429 [hep-th]].
%%CITATION = doi:10.1007/JHEP10(2014)178;%%
  %65 citations counted in INSPIRE as of 03 Dec 2018


%\cite{Camporesi:1990wm}
\bibitem{Camporesi:1990wm} 
  R.~Camporesi,
  %``Harmonic analysis and propagators on homogeneous spaces,''
Phys.\ Rept.\  {\bf 196}, 1 (1990).
doi:10.1016/0370-1573(90)90120-Q
%%CITATION = doi:10.1016/0370-1573(90)90120-Q;%%
  %218 citations counted in INSPIRE as of 03 Dec 2018


%\cite{Giombi:2008vd}
\bibitem{Giombi:2008vd} 
  S.~Giombi, A.~Maloney and X.~Yin,
  %``One-loop Partition Functions of 3D Gravity,''
JHEP {\bf 0808}, 007 (2008)
doi:10.1088/1126-6708/2008/08/007
[arXiv:0804.1773 [hep-th]].
%%CITATION = doi:10.1088/1126-6708/2008/08/007;%%
  %123 citations counted in INSPIRE as of 03 Dec 2018


%\cite{David:2009xg}
\bibitem{David:2009xg} 
  J.~R.~David, M.~R.~Gaberdiel and R.~Gopakumar,
  %``The Heat Kernel on AdS(3) and its Applications,''
JHEP {\bf 1004}, 125 (2010)
doi:10.1007/JHEP04(2010)125
[arXiv:0911.5085 [hep-th]].
%%CITATION = doi:10.1007/JHEP04(2010)125;%%
  %73 citations counted in INSPIRE as of 03 Dec 2018


%\cite{DeWitt:1965jb}
\bibitem{DeWitt:1965jb} 
  B.~S.~DeWitt,
  %``Dynamical theory of groups and fields,''
Conf.\ Proc.\ C {\bf 630701}, 585 (1964)
[Les Houches Lect.\ Notes {\bf 13}, 585 (1964)].
%%CITATION = CONFP,C630701,585;%%
  %146 citations counted in INSPIRE as of 03 Dec 2018


%\cite{Allen:1985ux}
\bibitem{Allen:1985ux} 
  B.~Allen,
  %``Vacuum States in de Sitter Space,''
Phys.\ Rev.\ D {\bf 32}, 3136 (1985).
doi:10.1103/PhysRevD.32.3136
%%CITATION = doi:10.1103/PhysRevD.32.3136;%%
  %508 citations counted in INSPIRE as of 03 Dec 2018


%\cite{Allen:1987tz}
\bibitem{Allen:1987tz} 
  B.~Allen and A.~Folacci,
  %``The Massless Minimally Coupled Scalar Field in De Sitter Space,''
Phys.\ Rev.\ D {\bf 35}, 3771 (1987).
doi:10.1103/PhysRevD.35.3771
%%CITATION = doi:10.1103/PhysRevD.35.3771;%%
  %253 citations counted in INSPIRE as of 03 Dec 2018
%
%
\end{thebibliography}
\end{document}